\documentclass[a4paper]{IEEEtran}
\usepackage[T1]{fontenc}
\usepackage[latin9]{inputenc}
\usepackage{color}
\usepackage{float}
\usepackage{amsmath}
\usepackage{amsthm}
\usepackage{amssymb}
\usepackage{stackrel}
\usepackage{graphicx}
\usepackage{esint}

\makeatletter

\pdfpageheight\paperheight
\pdfpagewidth\paperwidth

  \theoremstyle{plain}
  \newtheorem{thm}{\protect\theoremname}
  \theoremstyle{remark}
  \newtheorem{rem}{\protect\remarkname}

\usepackage{xcolor}
\usepackage{pdfcolmk}
\usepackage{units}

\usepackage{epsfig}
\usepackage{float}
\usepackage{subfigure}
\usepackage{subfloat}
\usepackage{cite}

\usepackage{miniplot}

\allowdisplaybreaks

\PassOptionsToPackage{normalem}{ulem}
\usepackage{ulem}

  \pdfpageheight\paperheight
\pdfpagewidth\paperwidth

\makeatother

\providecommand{\remarkname}{Remark}
\providecommand{\theoremname}{Theorem}

\begin{document}

\title{Throughput of Infrastructure-based Cooperative Vehicular Networks}

\author{Jieqiong Chen, \IEEEmembership{Student Member, IEEE}, Guoqiang Mao,
\IEEEmembership{Senior Member, IEEE}, \\
 Changle Li, \IEEEmembership{Senior Member, IEEE}, Ammar Zafar, \IEEEmembership{Member, IEEE},
Albert Y. Zomaya, \IEEEmembership{Fellow, IEEE}}
\maketitle
\begin{abstract}
\textcolor{black}{In this paper, we provide detailed analysis of the
achievable throughput of infrastructure-based vehicular network with
a finite traffic density under a cooperative communication strategy,
which explores combined use of vehicle-to-infrastructure (V2I) communications,
vehicle-to-vehicle (V2V) communications, mobility of vehicles and
cooperations among vehicles and infrastructure to facilitate the data
transmission. A closed form expression of the achievable throughput
is obtained, which reveals the relationship between the achievable
throughput and its major performance-impacting parameters such as
distance between adjacent infrastructure points, radio ranges of infrastructure
and vehicles, transmission rates of V2I and V2V communications and
vehicular density. Numerical and simulation results show that the
proposed cooperative communication strategy significantly increases
the throughput of vehicular networks, compared with its non-cooperative
counterpart, even when the traffic density is low. Our results shed
insight on the optimum deployment of vehicular network infrastructure
and optimum design of cooperative communication strategies in vehicular
networks to maximize the throughput.}
\end{abstract}

\begin{IEEEkeywords}
Data dissemination, cooperative communication, throughput, vehicular
networks, vehicle-to-infrastructure communications, vehicle-to-vehicle
communications. 
\end{IEEEkeywords}

\section{Introduction}

Vehicular networks have recently gained significant interest from
academia and industry because of their increasingly important role
in improving road traffic efficiency, enhancing road safety and providing
real-time information to drivers and passengers \cite{Li2014,S.AL2014,Ning14}.
By disseminating real-time information about traffic accidents, traffic
congestion or obstacles in the road, road safety and traffic efficiency
can be greatly improved. Furthermore, offering value-added services
like digital maps with real-time traffic status and in-car entertainment
services can greatly enhance the convenience and comfort of drivers
and passengers.

\textcolor{black}{Vehicle-to-infrastructure (V2I) and vehicle-to-vehicle
(V2V) communications, on one hand, are two fundamental techniques
to disseminate data for vehicular applications; on the other hand,
as pointed out in our previous paper \cite{Jchen16} and other work
\cite{Khabazian13,Reis14}, V2V communications may become unreliable
when the number of hops in the communication becomes large, and incur
long communication delay when the vehicular density is low. Moreover,
V2I communications may have limited availability due to the limited
number of infrastructure points attributable to the high deployment
cost, especially in rural areas and in the initial deployment phase
of vehicular network. Therefore, V2I and V2V communications may have
to co-exist and complement each other to meet the diverse communication
requirements of vehicular networks ranging from safety information
dissemination to in-car entertainment services.}

\textcolor{black}{In this paper, we consider a scenario where there
is a vehicle of interest (VoI) wanting to download a large-size file,
e.g., a video, from the Internet, and all the other vehicles (termed
}\textit{\textcolor{black}{helpers}}\textcolor{black}{) assist its
download using a cooperative communication strategy, which explores
the combined use of V2I communication, V2V communication, the mobility
of vehicles and cooperation among vehicles and infrastructure to facilitate
data transmission. The scenario being considered corresponds to the
category of delay-tolerant applications. We are interested in the
long-term data rate the VoI can achieve, i.e., the achievable throughput,
which is one of the most important performance metrics in wireless
(vehicular) networks because it characterizes the feasible data dissemination
rate of the network. In our previous work \cite{Jchen16}, under the
same network setting, we have analyzed the achievable throughput of
vehicular network with the assumption that the data rate of V2I communications
is larger than the data rate of V2V communications. In this paper,
we extend to consider a more general scenario without the restriction
of the aforementioned assumption and give an accurate analysis on
the achievable throughput by the VoI when there is finite vehicular
density in the network, and investigate the topological impact on
the achievable throughput. Our analytical results shed insight on
the optimum deployment of vehicular network infrastructure and the
design of optimum cooperative communication strategies in finite vehicular
networks to maximize the throughput.}

The novelty and major contributions of this paper are summarized as
follows: 
\begin{enumerate}
\item a cooperative communication strategy is proposed, which utilizes V2I
communications, V2V communications, the mobility of vehicles, and
cooperations among vehicles and infrastructure to improve the achievable
throughput by the VoI; 
\item an analytical framework is developed for studying the data dissemination
process under our cooperative communication strategy. A closed-form
expression of the achievable throughput by the VoI in a vehicular
network with a moderate number of vehicles or a finite vehicular density
is obtained, which reveals the relationship between the achievable
throughput and different parameters such as distance between two neighboring
infrastructure points, radio ranges of infrastructure and vehicles,
transmission rates of V2I communications and V2V communications and
density of vehicles;
\item both simulation and numerical results show that the proposed cooperative
communication strategy significantly increases the throughput of vehicular
networks, compared with its non-cooperative counterpart, even when
the traffic density is low. 
\end{enumerate}
The rest of this paper is organized as follows: Section \ref{sec:Related-Work}
reviews related work. Section \ref{sec:System-Model-and} introduces
the system model, the proposed cooperative communication strategy
and the problem formation. Theoretical analysis of the data dissemination
process and the achievable throughput are provided in Section \ref{sec:Theoretical-Analysis-of-V2V}.
In Section \ref{sec:Simulation-and-Discussion}, we validate the analytical
results using simulations and discuss the impact of major performance-impacting
parameters. Section \ref{sec:Conclusion-and-Future} concludes this
paper.

\section{Related Work\label{sec:Related-Work}}

Since the seminal work of Gupta and Kumar \cite{Gupta}, extensive
research has been done to investigate the throughput and capacity
of wireless networks. Particularly, Gupta and Kumar showed that the
maximum throughput of static wireless networks is $\Theta(\frac{1}{\sqrt{n}})$
with $n$ being the number of nodes in the network. In \cite{Gross02},
Grossglauser and Tse showed that by leveraging on the nodes' mobility,
a per-node throughput of $\Theta(1)$ can be achieved at the expense
of unbounded delay. In \cite{Mao13}, Mao et al. presented a simple
relationship to estimate the capacity of both static and mobile networks
and developed a generic methodology for analyzing the network capacity
that is applicable to a variety of different scenarios. Focusing on
the capacity of vehicular networks, Wang et al. \cite{Wang14} analyzed
the asymptotic uplink throughput scaling law of urban vehicular networks
with uniformly distributed RSUs.\textcolor{black}{{} All of the aforementioned
work focused on studying the scaling law of the throughput (capacity)
when the number of vehicles or vehicular density is sufficiently large.
In this paper, we focus on an accurate analysis, instead of the scaling
law, of the throughput of vehicular networks with a moderate number
of vehicles or a finite vehicular density where the asymptotic analysis
may not apply.}

Other work has also investigated the performance of vehicular networks,
measured by the downloaded data volume \cite{Haibo14}, transmission
delay \cite{Zhu15}, communication link quality \cite{ZhuYi15} etc.
Among the major techniques to enhance these performance measures,
cooperative communications, including both cooperations among vehicles
and cooperations among infrastructure points, stands out as a popular
and important technique. The following work has investigated cooperative
communications among vehicles in vehicular networks. In \cite{Haibo14},
Zhou et al. introduced a cooperative communication strategy using
a cluster of vehicles on the highway to cooperatively download the
same file from the infrastructure to enhance the probability of successful
download. In \cite{Zhu15}, Zhu et al. studied using multiple nearby
vehicles to collaboratively download data from an RSU and analyzed
the average download time using the network coding techniques. In
\cite{Yan12}, Yan et al. developed a theoretical model to analyze
the achievable channel data rate of VANETs for cooperative mobile
content dissemination, also assisted by network coding techniques.
\textcolor{black}{They focused on the transmission throughput in MAC
layer, which is different from this work as we focus on the network
achievable throughput.} Li et al. \cite{Li2011} proposed a push-based
popular content distribution scheme for vehicular networks, where
large files are broadcast proactively from a few infrastructure points
to vehicles inside an interested area and further disseminated cooperatively
among vehicles using V2V communications. In \cite{Wang2013}, Wang
et al. introduced a coalitional graph game to model the cooperations
among vehicles and proposed a coalition formation algorithm to implement
the cooperation between vehicles for popular content distribution.
In addition to cooperations among vehicles, cooperations among infrastructure
points can also be achieved by caching different files or different
parts of a file in different infrastructure points to help moving
vehicles to download from the Internet. In \cite{Z13}, to fully utilize
the wireless bandwidth provided by APs, Zhang and Yeo proposed a cooperative
content distribution system for vehicles by using cooperative APs
to distribute contents to moving vehicles. More specifically, by prefetching
different data into the selected APs, vehicles can obtain the complete
data from those selected APs when traveling through their coverage
areas. In \cite{Li2014} and \cite{Kim15}, by utilizing infrastructure
cooperation for data dissemination, the authors proposed a cooperative
content dissemination scheme in vehicular networks to maximize the
downloaded data size \cite{Li2014} and the success probability of
download \cite{Kim15} respectively. However, the aforementioned work
only considered either cooperations among vehicles or cooperations
among infrastructure points. In contrast, our work considers both
types of cooperations to maximize the throughput.

There are very limited studies considering both vehicular cooperation
and infrastructure cooperation. In \cite{Mershad12}, Mershad et al.
explored cooperations among inter-connected infrastructure and V2V
communications to efficiently deliver packets from a source vehicle
to vehicles far away. In their work, they focused on designing the
optimum routing path to reduce the total delay for delivering a packet
from a source to its destination. Our focus is on investigating the
achievable throughput in vehicular networks when a VoI requests data
from the Internet.

\section{System Model and Problem Formation\label{sec:System-Model-and}}

\subsection{Network Model}

We consider a highway with bi-directional traffic flows. The highway
is modeled by an infinite line with roadside infrastructure, e.g.,
RSUs, Wi-Fi APs or LTE base stations, regularly deployed with equal
distance $d$. The width of a lane is typically small compared with
the transmission range of vehicles. Therefore, we ignore the road
width and model multiple lanes in the same direction as one lane \cite{Abboud14,Zhang12,Wis2007}.
We further assume that all infrastructure points are connected to
the Internet through wired or wireless infrastructure.

\begin{figure}
\centering{}\includegraphics[scale=0.8]{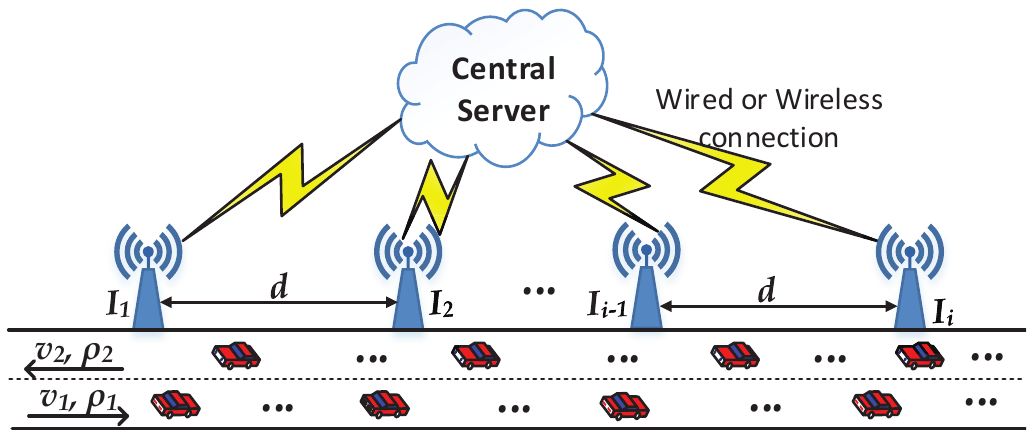} \caption{An illustration of the system model for a bi-directional highway with
infrastructure regularly deployed with equal distance $d$. The density
and speed of vehicles in each direction are $\rho_{1}$, $v_{1}$
and $\rho_{2}$, $v_{2}$ respectively.\label{Fig: System model} }
\end{figure}

\textcolor{black}{We adopt a widely used traffic model in highway
\cite{Reis14,Wis2007,Ruixue13}, that in each direction (eastbound
and westbound), the distribution of vehicles follows a homogeneous
Poisson process with densities $\rho_{1}$ and $\rho_{2}$ respectively.
It follows that the inter-vehicle distances in each direction are
exponentially distributed. This exponential inter-vehicle spacing
distribution has been supported by some empirical study that it can
accurately characterize the real traffic distribution when the traffic
density is low or medium \cite{Wis2007}}. Besides, vehicles in each
direction travel at the same constant speed of $v_{1}$ and $v_{2}$
respectively \cite{Reis14,Wis2007,Wei2014}. In real networks, individual
vehicular speed may deviate from the mean speed, e.g., Gaussian speed
model \cite{Zhang2014,Ge15}. However, such deviations, which results
in vehicle overtaking, have only minor impact on the throughput being
studied as shown later in our simulation. The system model is illustrated
in Fig. \ref{Fig: System model}.

\subsection{Wireless Communication Model\label{subsec:Wireless-Communication-Model}}

Both V2I and V2V communications are considered. All infrastructure
points have the same radio range, denoted by $r_{I}$; and all vehicles
have the same radio range, denoted by $r_{0}$ where $r_{I}>r_{0}$,
which reflects the fact that infrastructure typically has stronger
communication capability. A pair of vehicles (or vehicle and infrastructure)
can directly communicate with each other if and only if their Euclidean
distance is not larger than the radio range $r_{0}$ (or $r_{I}$).
\textcolor{black}{There are other more realistic and intricate connection
models, e.g., the SINR connection model \cite{Du2015} and the log-normal
connection model \cite{ZijieICC14,Mao2006}. This simple unit disk
model has been extensively used in the field \cite{Reis14,Zhang12,Wee11,Mao09}.
It grossly captures the fact that all wireless devices have a limited
transmission range and that the closer two devices become, the easier
it is for them to establish a connection. This simplification allows
us to omit physical layer details and focus on the topological impact
of vehicular networks on the throughput, which is the main performance
determining factor. We will show later in the simulation that the
unit disk model assumption has little impact on the throughput.}

We consider that each vehicle has a single antenna so that they can
not transmit and receive at the same time. Besides, we adopt unicast
transmission that each infrastructure (or vehicle) can only transmit
information to one vehicle at a time. \textcolor{black}{Note that
in many applications, e.g., sharing emergency or road traffic information
among multiple vehicles, broadcast is better used. The scenario being
considered in this paper corresponds to a unicast scenario where each
user may receive distinct content from the Internet. Both unicast
and broadcast are important in vehicular networks \cite{Cheng15}.
Furthermore, it has been shown in \cite{Gupta} that whether the infrastructure
transmit to one vehicle at a time, or divides its bandwidth among
multiple users and transmits to multiple users at the same time, does
not affect throughput calculation. Therefore, the assumption that
the infrastructure transmits to one vehicle at a time is immaterial
to throughput calculation. For the interference model, we assume that
V2I and V2V communications are allocated different channels so that
there is no mutual interference between them. Besides, we adopt the
widely used Protocol Interference Model \cite{Kumar04} that a transmitter
cannot transmit if there are other transmitter transmitting within
its interference range. In our work, the inter-infrastructure distance
is large so that infrastructure can transmit simultaneously without
causing any mutual interference. For the V2V communications, noting
that the VoI is the only receiver of V2V communications, therefore,
under the Protocol Interference Model, there will at most one transmitter
transmitting its data to the VoI at a randomly chosen time instant,
which implies that there will be no interference caused by other simultaneous
transmitters. Our work assuming the Protocol Interference Model can
be readily extended to another widely used Physical Interference Model
(also known as the SINR Model) because it has been established in
\cite{Kumar04} that any spatio-temporal scheduling satisfying the
Protocol Interference Model can also meet the requirement of the corresponding
Physical Interference Model when some parameters, like the interference
range, transmit power and the SINR threshold are appropriate selected.
Furthermore, the MAC protocol associated with the Protocol Interference
Model is the CSMA scheme. As we consider a single VoI only in this
paper, collisions, which are major concerns of MAC protocol, have
little impact on the achievable throughput of the VoI.}

We assume V2I and V2V communicate at a constant data rate $w_{I}$
and $w_{V}$ respectively \cite{Gupta,Gross02,Mao13}. For time-varying
channels, the values of $w_{I}$ and $w_{V}$ can be replaced by the
respective time-averaged data rate of V2I and V2V communications and
our analysis still applies. Furthermore, differently from our previous
paper \cite{Jchen16} which assume that $w_{I}>w_{V}$, in this paper,
we remove this assumption and give detailed analysis covering a wider
range of scenarios. Indeed, analysis later in this paper will show
that depending on the relationship between $w_{I}$, $w_{V}$ and
the speeds of vehicles in both directions, the system can be classified
into three regimes: one regime where the throughput is mainly limited
by the data rate of V2I communications, i.e., $w_{I}$; one regime
where the throughput is mainly restrained by the data rate of V2V
communications, $w_{V}$; and another regime where the throughput
is determined both by the data rate of V2I communications, $w_{I}$,
and the data rate of V2V communications, $w_{V}$. Furthermore, only
one-hop communications\textit{ }are considered.\textcolor{black}{{}
This can be explained by the fact that in the specific scenario being
considered, there is only one vehicle with download request (VoI),
all other vehicles (}\textit{\textcolor{black}{helpers}}\textcolor{black}{)
assist the VoI to receive more data. Any new data in the vehicular
network must come from the infrastructure. Therefore, allowing multi-hop
V2V communications between the VoI and }\textit{\textcolor{black}{helpers}}\textcolor{black}{,
e.g., allowing V2V communications between }\textit{\textcolor{black}{helpers}}\textcolor{black}{,
only helps to balance the distribution of information among }\textit{\textcolor{black}{helpers}}\textcolor{black}{{}
but do not increase the net amount of information available in the
network. Furthermore, even though allowing more than one hop communication
between the VoI and infrastructure is beneficial to the VoI's data
downloading because it allows the VoI having longer connection time
with the infrastructure, the improvement is expected to be marginal,
especially when the traffic density is small, which has been verified
by our simulation result as shown later.}

\subsection{Cooperative Communication Strategy}

Now we introduce the cooperative communication strategy considered
in this paper. Specially, we consider a scenario where a VoI wants
to download a large file, e.g., a video, from a remote server, and
analyze the throughput that can be achieved by the VoI via a combined
use of V2I communications, V2V communications, vehicular mobility
and cooperations among vehicles and infrastructure. 

The scenario being considered corresponds to a vehicular network where
only a small number of vehicles have requests for large-file downloads.
Another scenario that has been widely considered in the literature,
commonly known as \emph{the saturated traffic scenario}, considers
that all vehicles have requests for download. Saturated traffic scenario
is often used in analyzing the capacity of the network \cite{Gross02},\cite{Mao13}.
We point out that for the particular problem being considered, i.e.,
downloading a large file from a remote server, saturated traffic scenario
constitutes a trivial case and offer the following intuitive explanation
for that. Note that when downloading files from a remote server, the
new information (e.g., parts of the files) must come from the infrastructure
points. V2V communications only help to balance the distribution of
information among vehicles and do not increase the net amount of information
available in the system. Therefore, when all vehicles have download
requests, it can be easily established that the optimum strategy that
maximizes the capacity is for each vehicle to download its requested
file directly from the infrastructure. When only single vehicle or
a small number of vehicles have download requests, the situation becomes
more intriguing. In this situation, other vehicles may help to retrieve
information from the infrastructure when these vehicles enter into
the coverage of their respective infrastructure points and then deliver
the information to the VoI(s) outside the coverage of the infrastructure.
In this way, the net amount of new information available in the system
is boosted and therefore increasing the throughput (capacity) of the
VoI(s).

As mentioned in the beginning of this subsection, the VoI wants to
download a large file from the remote server. This requested large
file may be first split into multiple pieces and transmitted to different
infrastructure points so that each infrastructure point has a different
piece of data, which enables cooperation among infrastructure. Data
delivered to an infrastructure point may be further split and transmitted
to either the VoI or \textit{helpers} when they move into its coverage
so that \textit{helpers} have different pieces of data from each other
and from the VoI. Data received by the \textit{helpers} will be transmitted
to the VoI when they encounter the VoI, which exploits the mobility
of vehicles and V2V communications to achieves the vehicular cooperations.
Since vehicles in the same direction move at the same constant speed,
the inter-vehicle distances in the same lane remain the same at any
time instant. This follows that vehicles in the same direction of
the VoI can only offer limited help to the VoI because only vehicles
within the coverage of the VoI can offer help. Therefore, in this
paper, we only consider vehicles in the opposite direction\textit{
}of the VoI that will receive data from infrastructure and can transmit
the received data to the VoI as \textit{helpers.} 

To present the cooperative communication strategy, it suffices to
consider two consecutive infrastructure points along the travel direction
of the VoI collaborating to deliver data. Denote the nearest infrastructure
point along the travel direction of the VoI by $I_{1}$ and the second
nearest one by $I_{2}$. When the VoI is in the coverage of $I_{1}$,
it receives data directly from $I_{1}$. In the meantime, the \textit{helpers}\emph{
}may also receive different pieces of data from $I_{2}$ when they
move through the coverage of $I_{2}$. When the VoI moves outside
the coverage of $I_{1}$, it may continue to receive data from \textit{helpers}
using V2V communications. Of course, when the VoI moves along its
direction, the two infrastructure points participating in the cooperative
communication are also updated. In this way, V2I communications between
the VoI and infrastructure, between \textit{helpers} and their respective
infrastructure points, V2V communications between the VoI and \textit{helpers},
as well as vehicular mobility are coherently combined to maximize
the throughput of the VoI.\textcolor{black}{{} In our considered network
scenario, V2I communications by both the VoI and }\textit{\textcolor{black}{helpers}}\textcolor{black}{{}
are essential to retrieve data from the infrastructure. V2V communications
only help to assist the VoI to retrieve more data from the Internet
and deliver the data retrieved by }\textit{\textcolor{black}{helpers}}\textcolor{black}{{}
to the VoI. V2V communications can not increase the net amount of
data in the network. }Furthermore, we consider that some practical
issues like out of sequence data delivery can be handled by techniques
such as network coding (e.g., our recent paper \cite{WangPeng15})
so that we can focus on the main theme of the paper without the need
for considering their impacts.

\subsection{Problem Formation\label{subsec:Problem-Formation}}

Now we give a formal definition of the throughput studied in this
paper. Consider an arbitrarily chosen time interval $[0,t]$ and denote
the amount of data received by the VoI as $D(t)$, which includes
data received from both infrastructure and \textit{helpers}. In this
paper, we are interested in finding the long-term achievable throughput
of the VoI, using our cooperative communication strategy, where the
long-term throughput, denoted by $\eta$, is formally defined as follows:
\begin{equation}
\eta=\underset{t\rightarrow\infty}{\lim}\frac{D(t)}{t}.\label{eq:definition of throughput}
\end{equation}

Without loss of generality, we assume that the VoI travels at speed
$v_{1}$, and the \textit{helpers} travel at speed $v_{2}$ and have
vehicular density $\rho_{2}$. We define the time interval starting
from the time instant when the VoI enters into the coverage of one
infrastructure point to the time instant when the VoI enters into
the coverage of the next infrastructure point as \emph{one cycle}.
By using \textit{cycles} as the basic blocks, the entire data receiving
process of the VoI can be modeled by a renewal reward process \cite{Gallager13}.
Each \textit{cycle} in the renewal reward process consists of one
V2I communication process, followed by a V2V communication process,
and the reward is the amount of data received by the VoI during each
\textit{cycle}. This follows that the throughput can be calculated
as follows:
\begin{equation}
\eta=\underset{t\rightarrow\infty}{\lim}\frac{D(t)}{t}=\frac{E[D_{I}]+E[D_{V}]}{E[T]},\label{eq: alternative expression of throughput}
\end{equation}
where $E[D_{I}]$ and $E[D_{V}]$ are respectively the expected amount\textcolor{blue}{{}
}of data received by the VoI during the V2I communication process
and the V2V communication process in one \textit{cycle}, and $E[T]$
is the expected time of one \textit{cycle}, which can be calculated
as $E[T]=\frac{d}{v_{1}}$. Since when the VoI is covered by an infrastructure
point, it will only use V2I communication, $E[D_{I}]$ can be readily
obtained as follows: 
\begin{equation}
E[D_{I}]=\frac{2r_{I}w_{I}}{v_{1}}.\label{eq: DI}
\end{equation}

Using \eqref{eq: alternative expression of throughput} and \eqref{eq: DI},
the problem of calculating the achievable throughput by the VoI can
be transformed into the problem of calculating the expected amount
of data that can be received by the VoI from V2V communications in
one \textit{cycle}. Without loss of generality, we call the two infrastructure
points $I_{1}$ and $I_{2}$ respectively as defined earlier\textit{.}
Because of the unicast transmission model we adopt, during V2V communications
between the VoI and \textit{helpers}, the amount of data two adjacent
\textit{helpers}\emph{ }can deliver to the VoI become correlated when
their inter-vehicle distance is smaller than $2r_{0}$ and is further
limited by the amount of data each \textit{helper} receives from $I_{2}$,
which can also be correlated because during \textit{helpers'}\emph{
}V2I communications, the amount of data received by adjacent \textit{helpers}
become correlated when their inter-vehicle distance is smaller than
$2r_{I}$. This complicated correlation structure is quite intricate
for statistical analysis. In this paper, we handle the challenge by
formulating the V2V data delivering process in one \textit{cycle}
as a constrained optimization problem, with the goal of obtaining
the maximum amount of data received by the VoI from \textit{helpers}
and finding the corresponding scheduling scheme, which includes V2I
transmission scheme for \textit{helpers} and V2V transmission scheme,
to reach this maximum value. In the following, we will show the formation
of the constrained optimization problem\textit{.} 

\begin{figure}
\centering{}\includegraphics[scale=1.9]{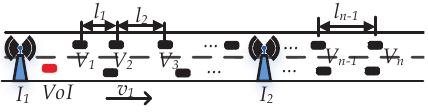} \caption{An illustration of \textit{helpers} encountered by the VoI during
one V2V communication \textit{cycle} and their interval distance.
\label{V2V} }
\end{figure}

Denote by $n$ the number of \textit{helpers}\emph{ }encountered by
the VoI during a \textit{cycle} and $n$ is a Poissonly distributed
random integer. Denote by $V_{1}$ the first \textit{helper}  encountered
by the VoI when the VoI moves outside the coverage of $I_{1}$, by
$V_{2}$ the second \textit{helper}, and so on. Denote the distance
between two consecutive \textit{helpers} $V_{i}$ and $V_{i+1}$ by
$l_{i},i=1,...n-1$. See Fig. \ref{V2V} for an illustration. Furthermore,
denote by $D_{i}$ the amount of data received by \textit{helper}
$V_{i}$ from $I_{2}$, and by $Y_{i}$ the amount of data delivered
by $V_{i}$ to the VoI. We first consider the situation that $n$
is a fixed integer and $l_{i},i=1,...n-1$ are known values, i.e.,
corresponding to a specific instance of these random values, and then
extend to consider the more general situation that $n$ and $l_{i},i=1,...n-1$
are random values.

Without considering the boundary case, caused by \textit{helpers}
located near the borders of the coverage area of infrastructure points,
the problem of finding the maximum amount of data received by the
VoI from V2V communications in one \textit{cycle}, given $n$ and
$l_{i},i=1,...n-1$, can be formulated as the following optimization
problem. We will show later in the simulation that the boundary case
has negligible impact on the achievable throughput as we are focusing
on the long-term throughput. The optimization is taken over the set
of all possible scheduling schemes. Without causing any confusion,
we drop the notation for the set of all possible scheduling schemes
to have a simpler expression.
\begin{align}
\max & \sum_{i=1}^{n}Y_{i}\label{eq:formulation of the optimization problem}\\
s.t. & 0\leq D_{i}\leq\frac{2r_{I}}{v_{2}}w_{I,}\;i=1,2,...n\label{eq:constraint on Di}\\
 & \sum_{i=k_{1}}^{k_{2}}D_{i}\leq\frac{\sum_{i=k_{1}}^{k_{2}-1}\min\left\{ l_{i},2r_{I}\right\} +2r_{I}}{v_{2}}w_{I},\nonumber \\
 & \;\;\;\;\;\;\;\;\;\;\;\;\;\;\;\;\;\;\;\;\;\;\;\;\;\;\;\;\;\;\;\;\;\;\;\;\;\;\;\;\;\;\;1\leq k_{1}\leq k_{2}\leq n\label{eq:constraint on sum of Di}\\
 & 0\leq Y_{i}\leq\frac{2r_{0}}{v_{1}+v_{2}}w_{V},\text{\,\,\,\,}i=1,2,...n\label{eq:constraint on Yi}\\
 & Y_{i}\leq D_{i},\text{\,\,\,\,}i=1,2,...n\label{eq:Yi<Di}\\
 & \sum_{i=k_{1}}^{k_{2}}Y_{i}\leq\frac{\sum_{i=k_{1}}^{k_{2}-1}\min\left\{ l_{i},2r_{0}\right\} +2r_{0}}{v_{1}+v_{2}}w_{V},\nonumber \\
 & \;\;\;\;\;\;\;\;\;\;\;\;\;\;\;\;\;\;\;\;\;\;\;\;\;\;\;\;\;\;\;\;\;\;\;\;\;\;\;\;\;\;\;1\leq k_{1}\leq k_{2}\leq n\label{eq:constraint on sum of Yi}
\end{align}

In the above optimization problem, $\sum_{i=1}^{n}Y_{i}$ is the total
amount of data received by the VoI from \textit{helpers} during one
\textit{cycle}. Constraint \eqref{eq:constraint on Di} gives the
maximum and the minimum amount of data received by each \textit{helper}
from infrastructure point $I_{2}$. Constraint \eqref{eq:constraint on sum of Di}
gives an upper bound on the amount of data that any $k_{2}-k_{1}+1$
consecutive \textit{helpers} can receive from $I_{2}$, where the
term $\frac{\sum_{i=k_{1}}^{k_{2}-1}\min\left\{ l_{i},2r_{I}\right\} +2r_{I}}{v_{2}}$
gives the total amount of time these $k_{2}-k_{1}+1$ consecutive
\textit{helpers}\emph{ }can receive data from $I_{2}$. Particularly,
due to the randomness of vehicle distributions, it may happen that
there exists a void region of larger than $2r_{I}$, which has no
vehicle (\textit{helper}). When the void region occurs, \textit{helpers}\emph{
}may not be able to receive data continuously from $I_{2}$. Therefore,
all two constraints \eqref{eq:constraint on Di} and \eqref{eq:constraint on sum of Di}
must be considered to completely describe the V2I communication between
$I_{2}$ and \textit{helpers}. Constraint \eqref{eq:constraint on sum of Di}
also captures the correlation that may occur during the data receiving
process of adjacent \textit{helpers}, which has been explained earlier.
Similarly, constraint \eqref{eq:constraint on Yi} gives the maximum
and minimum amount of data that can be received by the VoI from each
\textit{helper}. Constraint \eqref{eq:Yi<Di} implies that the amount
of data each \textit{helper}  can deliver to the VoI cannot exceed
the data it receives from $I_{2}$. Constraint \eqref{eq:constraint on sum of Yi}
gives the upper bound of the amount of data the VoI can receive from
any ${\color{blue}k_{2}-k_{1}+1}$ consecutive \textit{helpers}\emph{.}

\section{Theoretical Analysis of V2V Communication Process and Achievable
Throughput\label{sec:Theoretical-Analysis-of-V2V}}

\textcolor{black}{The data received by the VoI, whether through V2I
communications directly from the infrastructure or through V2V communications
from }\textit{\textcolor{black}{helpers}}\textcolor{black}{, eventually
comes from the infrastructure. }Intuitively, as we increase the data
rate of V2I communications, $w_{I}$, from a very small value while
keeping other parameters constant, initially the throughput will be
limited by the data rate of V2I communications. We call this regime
the \textit{Infrastructure-Limited Regime}\textit{\emph{. As we further
increase the value of $w_{I}$, we will reach a }}\textit{Transitional
Regime }\textit{\emph{where both the data rate of V2I communications
and the data rate of V2V communications play major roles in determining
the throughput of the VoI. If we increase the value of $w_{I}$ further
to a very large value, V2I communications will no longer be a bottleneck
in determining the throughput of the VoI. Instead, the data rate of
V2V communication, $w_{V}$, becomes the determining factor of the
VoI throughput. We call this regime the }}\textit{V2V-Limited Regime}\textit{\emph{.}} 

It is evident from the optimization problem \eqref{eq:formulation of the optimization problem}
that the amount of data received from V2V communications by the VoI
given fixed $n$ and $l_{i},i=1,...n$, $\sum_{i=1}^{n}Y_{i}$, satisfies:
\begin{equation}
\sum_{i=1}^{n}Y_{i}\leq\min\{D_{Vu1},D_{Vu2}\}\label{eq:DV<min DV1 DV2}
\end{equation}
where $D_{Vu1}=\frac{\sum_{i=1}^{n-1}\min\{l_{i},2r_{I}\}+2r_{I}}{v_{2}}w_{I}$
comes from a combination of constraints \eqref{eq:constraint on sum of Di}
and \eqref{eq:Yi<Di}, representing the maximum amount of data all
\textit{helpers} can receive from infrastructure; and $D_{Vu2}=\frac{\sum_{i=1}^{n-1}\min\{l_{i},2r_{0}\}+2r_{0}}{v_{1}+v_{2}}w_{V}$
comes from constraint \eqref{eq:constraint on sum of Yi}, representing
the maximum amount of data all \textit{helpers} can deliver to the
VoI through V2V communications without considering the limitation
of the amount of data they receive. When $0<w_{I}\leq\frac{r_{0}w_{V}v_{2}}{r_{I}(v_{1}+v_{2})}$,
we have $D_{Vu1}\leq D_{Vu2}$, which implies that the amount of data
the VoI can receive from \textit{helpers} is limited by the the amount
of data \textit{helpers} can receive through their\emph{ }V2I communications,
thus limited by $w_{I}$. Thus, we define the \textit{Infrastructure-Limited
Regime}: $0<w_{I}\leq\frac{r_{0}w_{V}v_{2}}{r_{I}(v_{1}+v_{2})}$.
Similarly, when $w_{I}\geq\frac{w_{V}v_{2}}{v_{1}+v_{2}}$, we have
$D_{Vu1}>D_{Vu2}$, which implies that the amount of data the VoI
can receive from \textit{helpers} is limited by the amount of data
the \textit{helpers} can deliver through V2V communications, thus
limited by $w_{V}$. Therefore, we define the \textit{V2V-Limited
Regime}: $w_{I}\geq\frac{w_{V}v_{2}}{v_{1}+v_{2}}$. The rest of the
region forms the \textit{Transitional Regime}: $\frac{r_{0}w_{V}v_{2}}{r_{I}(v_{1}+v_{2})}<w_{I}<\frac{w_{V}v_{2}}{v_{1}+v_{2}}$.

In the following subsections, we analyze the achievable throughput
by the VoI under each regime separately. 

\subsection{\textit{\emph{\label{subsec:Infrastructure-Limited-Regime:-}}}\textit{Infrastructure-Limited
Regime: }$0<w_{I}\leq\frac{r_{0}w_{V}v_{2}}{r_{I}(v_{1}+v_{2})}$
\label{subsec:Infrastructure-Limited-Regime}}

In this subsection, we first analyze the maximum amount of data that
can be received from \textit{helpers} by the VoI in one \textit{cycle}\emph{
}by solving the optimization problem \eqref{eq:formulation of the optimization problem},
and find the corresponding scheduling scheme to achieve this maximum
solution given fixed $n$ and $l_{i},i=1,...n-1$ . Then, we extend
to consider that $n$ and $l_{i},i=1,...n-1$ are random values, corresponding
to Poisson distribution of vehicles, and analyze the achievable throughput
under the obtained optimal scheduling scheme.

\subsubsection{An analysis of the V2V communication process }

The following theorem summarizes the major result of this subsection.
\begin{thm}
\label{thm:In-the-infrastructure-limited-maximum-data}In the Infrastructure-Limited
regime, given fixed $n$ and $l_{i},i=1,2,...n-1$, the maximum amount
of data the VoI can receive from all $n$ helpers in one cycle is
given by 
\begin{equation}
\left(\sum_{i=1}^{n}Y_{i}\right)_{1}^{*}=\frac{\sum_{i=1}^{n-1}\min\{l_{i},2r_{I}\}+2r_{I}}{v_{2}}w_{I}\label{eq: optimal DV in Infrastructure-Limited Regime}
\end{equation}
where $\left(\sum_{i=1}^{n}Y_{i}\right)_{1}^{*}$ is the respective
$\sum_{i=1}^{n}Y_{i}$ associated with its optimum value and we use
the subscript $1$ to mark the Infrastructure-Limited regime and superscript
$*$ to mark the optimum value.

Furthermore, there exists a V2I transmission scheme for helpers and
a V2V transmission scheme to reach the above maximum amount of received
data for the VoI, satisfying: 
\begin{equation}
\begin{cases}
Y_{1i}^{*}=D_{1i}^{*}=\frac{\min\{l_{i},2r_{I}\}}{v_{2}}w_{I}, & i=1,2,...n-1\\
Y_{1n}^{*}=D_{1n}^{*}=\frac{2r_{I}}{v_{2}}w_{I}
\end{cases}\label{eq: optimal trasnmission scheme in Infrastructure-Limited Regime}
\end{equation}
where $D_{1i}^{*}$ and $Y_{1i}^{*},i=1,...,n$ are the respective
$D_{i}$ and $Y_{i},i=1,...,n$ associated with the optimum solution. 
\end{thm}
\begin{IEEEproof}
In the Infrastructure-Limited regime, with conditions $w_{I}\leq\frac{r_{0}w_{V}v_{2}}{r_{I}(v_{1}+v_{2})}$
and $r_{I}>r_{0}$, we have $\frac{2r_{I}}{v_{2}}w_{I}\leq\frac{2r_{0}}{v_{1}+v_{2}}w_{V}$
and $\frac{\min\{l_{i},2r_{I}\}}{v_{2}}w_{I}\leq\frac{\min\{l_{i},2r_{0}\}}{v_{1}+v_{2}}w_{V},i=1,...n-1$.
It follows that: 
\begin{equation}
\frac{\sum_{i=1}^{n-1}\min\{l_{i},2r_{I}\}+2r_{I}}{v_{2}}w_{I}\leq\frac{\sum_{i=1}^{n-1}\min\{l_{i},2r_{0}\}+2r_{0}}{v_{1}+v_{2}}w_{V}
\end{equation}
The above equation implies that in the optimization problem \eqref{eq:formulation of the optimization problem},
constraint \eqref{eq:constraint on Yi} is redundant and constraint
\eqref{eq:constraint on sum of Yi} can be replaced with a tighter
constraint after merging the two constraints \eqref{eq:constraint on sum of Di}
and \eqref{eq:Yi<Di}. The new constraints for the optimization problem
\eqref{eq:formulation of the optimization problem} under Infrastructure-Limited
regime are shown as follows:
\begin{align}
 & 0\leq Y_{i}\leq D_{i}\leq\frac{2r_{I}}{v_{2}}w_{I},\text{\,\,\,\,}i=1,2,...n\label{eq:converted constraints on Yi -1}\\
 & \sum_{i=k_{1}}^{k_{2}}Y_{i}\leq\sum_{i=k_{1}}^{k_{2}}D_{i}\leq\frac{\sum_{i=k_{1}}^{k_{2}-1}\min\left\{ l_{i},2r_{I}\right\} +2r_{I}}{v_{2}}w_{I},\nonumber \\
 & \;\;\;\;\;\;\;\;\;\;\;\;\;\;\;\;\;\;\;\;\;\;\;\;\;\;\;\;\;\;\;\;\;\;\;\;\;\;\;\;1\leq k_{1}\leq k_{2}\leq n\label{eq: converted constraints on sum of Yi -1}
\end{align}

Constraint \eqref{eq: converted constraints on sum of Yi -1} shows
that an upper bound of $\sum_{i=1}^{n}Y_{i}$ is given by $\frac{\sum_{i=1}^{n-1}\min\{l_{i},2r_{I}\}+2r_{I}}{v_{2}}w_{I}$.
In the following, we will show that this upper bound is exactly the
optimum solution of $\sum_{i=1}^{n}Y_{i}$ in the Infrastructure-Limited
regime and can be reached under some scheduling scheme. 

Noting that the upper bound of $\sum_{i=1}^{n}Y_{i}$ , $\frac{\sum_{i=1}^{n-1}\min\{l_{i},2r_{I}\}+2r_{I}}{v_{2}}w_{I}$,
is the sum of $n$ separate components, with each component smaller
than or equal to $\frac{2r_{I}}{v_{2}}w_{I}$. Therefore, when each
$Y_{i}$ is equal to $D_{i}$, and is further equal to the corresponding
component forming $\frac{\sum_{i=1}^{n-1}\min\{l_{i},2r_{I}\}+2r_{I}}{v_{2}}w_{I}$,
i.e., when each $Y_{i}$ and each $D_{i},i=1,...n$ are given by \eqref{eq: optimal trasnmission scheme in Infrastructure-Limited Regime},
the value of $\sum_{i=1}^{n}Y_{i}$ will reach its upper bound $\frac{\sum_{i=1}^{n-1}\min\{l_{i},2r_{I}\}+2r_{I}}{v_{2}}w_{I}$
while satisfying the constraints in optimization problem \eqref{eq:formulation of the optimization problem}.
This leads to the expression of \eqref{eq: optimal DV in Infrastructure-Limited Regime}.

\textcolor{black}{It remains to demonstrate that there exists a scheduling
scheme to reach this optimum solution specified in \eqref{eq: optimal DV in Infrastructure-Limited Regime}.
To this end, we show that \eqref{eq: optimal trasnmission scheme in Infrastructure-Limited Regime}
readily leads to the design of an optimal transmission scheme. Specifically,
a scheduling scheme which schedules both V2I and V2V transmissions
on a first-in-first-out (FIFO) basis can achieve the optimum solution
\eqref{eq: optimal DV in Infrastructure-Limited Regime}. We acknowledge
that the optimum scheduling algorithm that achieves the optimum solution
\eqref{eq: optimal DV in Infrastructure-Limited Regime} may not be
unique. When other performance metric is considered, e.g., delay,
the earliest deadline first scheme may have better delay performance
while achieving the same throughput. Particularly, in the FIFO scheduling
scheme, each }\textit{\textcolor{black}{helper}}\textcolor{black}{{}
starts its V2I communication once it enters the coverage of infrastructure
point $I_{2}$ }\textcolor{black}{\emph{and}}\textcolor{black}{{} there
is no other }\textit{\textcolor{black}{helper}}\textcolor{black}{{}
preceding it communicating with the infrastructure point $I_{2}$,
and stops when the }\textit{\textcolor{black}{helper}}\textcolor{black}{{}
leaves the coverage of the infrastructure point $I_{2}$, which lead
to that each }\textit{\textcolor{black}{helper}}\textcolor{black}{{}
will receive an amount of data shown as each $D_{1i}^{*},i=1,2,...n$
in Eq. \eqref{eq: optimal trasnmission scheme in Infrastructure-Limited Regime}.
Similarly, for V2V communications, the VoI receives data from each
}\textit{\textcolor{black}{helper}}\textcolor{black}{{} one by one when
there exists at least one }\textit{\textcolor{black}{helper}}\textcolor{black}{{}
within its coverage on a FIFO basis. Once a }\textit{\textcolor{black}{helper}}\textcolor{black}{{}
starts to deliver its data to the VoI, it will stop until it has transmitted
all its data to the VoI }\textit{\textcolor{black}{or}}\textcolor{black}{{}
when it leaves the coverage of the VoI, which lead to that the data
the VoI receives from each }\textit{\textcolor{black}{helper}}\textcolor{black}{{}
is shown as each $Y_{1i}^{*},,i=1,2,...n$ in Eq. \eqref{eq: optimal trasnmission scheme in Infrastructure-Limited Regime}.
Noting that Eq. \eqref{eq: optimal trasnmission scheme in Infrastructure-Limited Regime}
leads to the optimum solution \eqref{eq: optimal DV in Infrastructure-Limited Regime},
therefore, it can be readily established that the aforementioned scheduling
scheme achieves the maximum amount of the received data for the VoI
specified in \eqref{eq: optimal DV in Infrastructure-Limited Regime}. }
\end{IEEEproof}

\begin{rem}
\label{rem:valid for any n}Note that \eqref{eq: optimal DV in Infrastructure-Limited Regime}
and the corresponding scheduling scheme that satisfies \eqref{eq: optimal trasnmission scheme in Infrastructure-Limited Regime}
are valid for any value of $n$ and the corresponding $l_{i},i=1,...n-1$. 
\end{rem}

\subsubsection{Throughput calculation\label{subsec:Throughput-calculation}}

On the basis of Theorem \ref{thm:In-the-infrastructure-limited-maximum-data},
we now analyze the achievable throughput by the VoI considering that
both $n$ and the corresponding $l_{i},i=1,...n-1$ are random values.
A brute force approach of computing the achievable throughput will
first consider that $n$ is a Poisson random variable, then conditioned
on each value of $n$ (noting that conditional on a specific instance
of $n$, \textit{helpers} become uniformly distributed and hence $l_{i}$s,
$i=1,...n-1$, become correlated), evaluate the joint distribution
of the random variables $\min\left\{ l_{i},2r_{I}\right\} $, $i=1,...n-1$,
and finally transform the conditional value into unconditional one
using the total probability theorem and the Poisson distribution of
$n$. This will result in a very complicated analysis. In the following,
we use simpler techniques by resorting to the concept of \emph{clusters},
defined shortly later, to analyze the achievable throughput. 

We designate the time instant when the VoI leaves the coverage of
$I_{1}$ as $t=0$ and define its moving direction as the positive
(right) direction of the coordinate system. Furthermore, we define
the point to the right of $I_{1}$ and at a distance $r_{I}-r_{0}$
to $I_{1}$ as the origin of the coordinate system. It follows from
the above that the time instant when the VoI enters into $I_{2}$'s
coverage will be $t_{1}=\frac{d-2r_{I}}{v_{1}}$. Noting that the
relative speed of the VoI to the \textit{helpers} traveling in the
opposite direction is $v_{1}+v_{2}$, therefore the relative distance
traveled by the VoI, relative to the \textit{helpers} in the opposite
direction which all travel at the same constant speed of $v_{2}$
during $\left[0,t_{1}\right]$, is given by $\frac{(d-2r_{I})}{v_{1}}(v_{1}+v_{2})$.
The random number of \textit{helpers} encountered by the VoI, who
may deliver data to the VoI, during $\left[0,t_{1}\right]$, is determined
by the parameter $s$: 
\begin{equation}
s=\frac{(d-2r_{I})(v_{1}+v_{2})}{v_{1}}+r_{0},
\end{equation}
where the $r_{0}$ term is due to the consideration that when the
VoI exits the coverage of $I_{1}$ and is located at coordinate $r_{0}$
(and at time instant $t=0$), the \textit{helper(s)} to the left of
the VoI and within a distance $r_{0}$ to the VoI may possibly deliver
data to the VoI too. Thus, all \textit{helpers} in the opposite direction
that the VoI may encounter during its V2V communication process in
one \textit{cycle} are within road segment $[0,s]$. 

\begin{figure}
\centering{}\includegraphics[scale=1.9]{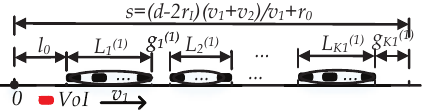} \caption{An illustration of \textit{clusters} formed by the \textit{helpers.}
Each \textit{cluster} has length $L_{j}^{(1)},j=1,...K_{1}$ and each
gap between two consecutive \textit{clusters} has length $g_{j}^{(1)},j=1,...K_{1}$.\label{fig: an illustration of cluster} }
\end{figure}

As explained in the beginning of this subsection, we use the concept
of \textit{clusters} to simplify our analysis. A \emph{cluster} is
defined as a maximal set of \textit{helpers} located within road segment
$\left[0,s\right]$ \emph{and }the distance between any two adjacent
\textit{helpers}  is smaller than or equal to $2r_{I}$. Forming \textit{clusters}
in this way allows us to remove the complexity associated with the
computation of the joint distribution of $\min\left\{ l_{i},2r_{I}\right\} ,i=1,...n-1$
because within each \textit{cluster}, we have $\min\left\{ l_{i},2r_{I}\right\} =l_{i},i=1,2,...$.
For each \textit{cluster}, we only need to focus on the length of
\textit{\emph{the}} \textit{cluster} rather than the individual inter-vehicle
distances. There may be multiple \textit{clusters} within road segment
$\left[0,s\right]$ and a \textit{cluster} may contain single vehicle
only. Denote the coordinate of the first \textit{helper} that can
transmit data to the VoI since $t=0$ by $l_{0}$. Due to the memoryless
property of the exponential distribution of inter-vehicle distances,
$l_{0}$ has the same exponential distribution as other $l_{i}$s,
$i=1,...n-1$ and the starting position of a vehicle in a \textit{cluster}
 does not affect the distribution of the length of the \textit{cluster}.
Denote by $K_{1}$ the random non-negative integer representing the
number of \textit{clusters} \emph{ }the VoI will encounter in one
\textit{cycle}. Furthermore, denote by $L_{j}^{(1)},j=1,...,K_{1}$,
the length of each \textit{clusters}\textit{\emph{,}} which are identically
and independently distributed (i.i.d), and by $g_{j}^{(1)},j=1,...K_{1}$
the length of each gap between two adjacent \textit{clusters}\textit{\emph{,}}
which are also i.i.d. See Fig. \ref{fig: an illustration of cluster}
for an illustration. 

Noting that \eqref{eq: optimal DV in Infrastructure-Limited Regime}
is also valid for any subset of \textit{helpers} within road segment
$\left[0,s\right]$ adopting the scheduling scheme described in the
proof of Theorem \ref{thm:In-the-infrastructure-limited-maximum-data},
therefore the amount of data each \textit{cluster}\emph{ }of \textit{helpers}
delivers to the VoI, denoted by $R_{j}^{(1)},j=1,...K_{1}$, can be
obtained as follows (recall that the analysis is conducted for the
Infrastructure-Limited regime): 
\begin{equation}
R_{j}^{(1)}=\frac{L_{j}^{(1)}+2r_{I}}{v_{2}}w_{I},j=1,...K_{1}\label{eq:data amount received from each cluster-1}
\end{equation}
It follows that the amount of data received by the VoI from \textit{helpers}
in one \textit{cycle}, denoted by $D_{V1}$, can be readily calculated
by summing up the amount of data received by the VoI from each \textit{cluster}\emph{
}of \textit{helpers}: 
\begin{equation}
D_{V1}=\sum_{j=1}^{K_{1}}R_{j}^{(1)}=\sum_{j=1}^{K_{1}}\frac{L_{j}^{(1)}+2r_{I}}{v_{2}}w_{I}.\label{eq:DV1 as expression of cluster length}
\end{equation}

Noting that in \eqref{eq:DV1 as expression of cluster length}, both
the number of \textit{clusters} in road segment $[0,s]$, $K_{1}$,
and the length of each \textit{cluster,} $L_{j}^{(1)}$, are random
variables, and they are not independent. If we \emph{approximately}
consider they are independent with each other, then from \eqref{eq:DV1 as expression of cluster length},
the expected amount of data received by the VoI from V2V communications
in one \textit{cycle}, $E[D_{V1}]$, can be calculated as follows
: 
\begin{equation}
E[D_{V1}]=E[K_{1}]\cdot\frac{E[L_{j}^{(1)}]+2r_{I}}{v_{2}}w_{I},\label{eq:expected value of DV1}
\end{equation}
where $E[K_{1}]$ is the expected number of \textit{clusters}  in
the road segment $[0,s]$.\textcolor{black}{{} The accuracy of this
}\textit{\textcolor{black}{approximation}}\textcolor{black}{{} is verified
by simulation, see Fig. \ref{EVAL-approx} below. It shows that the
}\textit{\textcolor{black}{approximation}}\textcolor{black}{{} will
marginally increase the result of $E[D_{V1}]$, which in turn, will
leads to a marginally increase in the achievable throughput in this
regime. }

\begin{figure}[H]
\begin{centering}
\includegraphics[width=8cm]{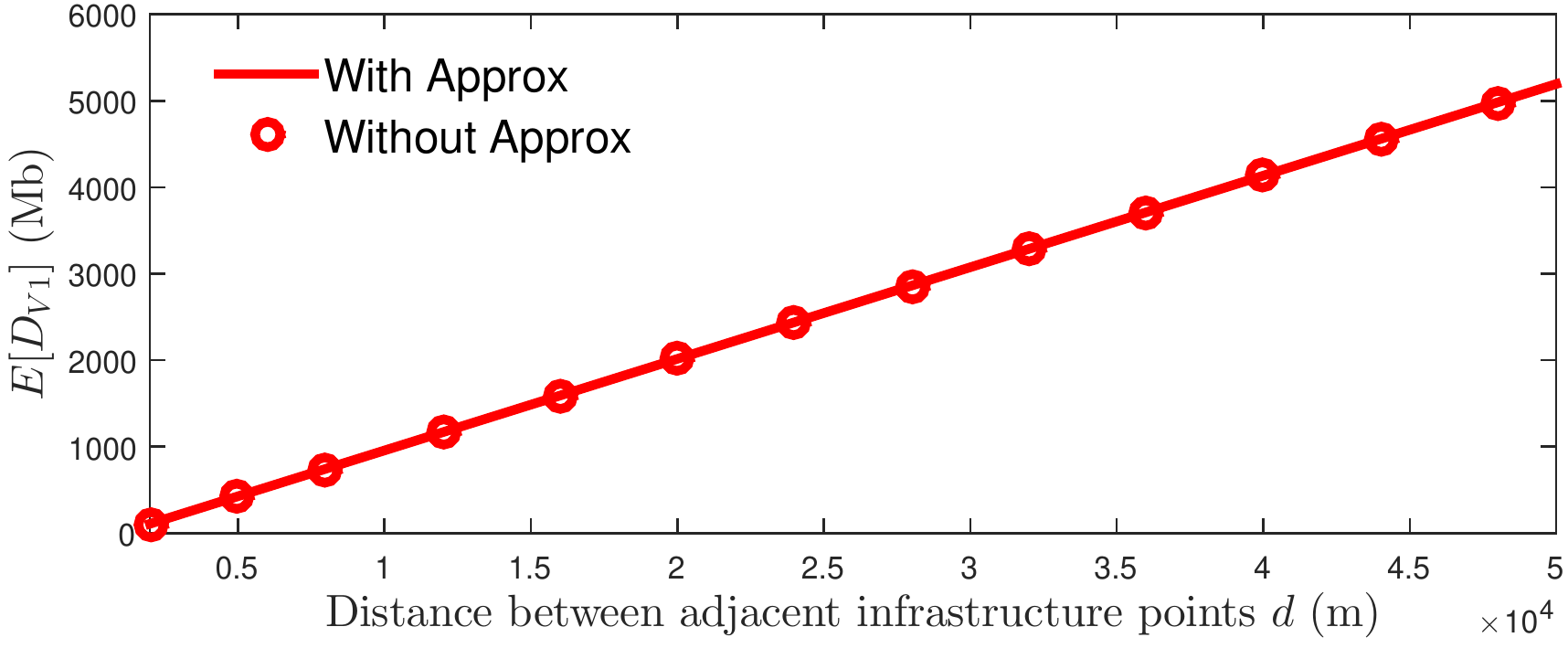}
\par\end{centering}
\caption{\textcolor{black}{A comparison between the result of $E[D_{V1}]$
with and without the }\textit{\textcolor{black}{approximation}}\textcolor{black}{.}}

\label{EVAL-approx}
\end{figure}

As $L_{j}^{(1)},j=1,...K_{1}$ and $g_{j}^{(1)},j=1,...K_{1}$ are
both i.i.d and each $L_{j}^{(1)}$ and $g_{j}^{(1)}$ are also mutually
independent, then according to the Generalized Wald's equality \cite[Theorem 4.5.2]{Gallager13},
when $s\gg E[L_{j}^{(1)}]+E[g_{j}^{(1)}]$, $E[K_{1}]$ can be approximately
calculated as follows: 
\begin{equation}
E[K_{1}]=\frac{s-E[l_{0}]}{E[L_{j}^{(1)}]+E[g_{j}^{(1)}]}.\label{eq:EK in infrastructure-limited regime}
\end{equation}
By putting \eqref{eq:EK in infrastructure-limited regime} into \eqref{eq:expected value of DV1},
we have 
\begin{equation}
E[D_{V1}]=\frac{s-E[l_{0}]}{E[L_{j}^{(1)}]+E[g_{j}^{(1)}]}\times\frac{E[L_{j}^{(1)}]+2r_{I}}{v_{2}}w_{I},\label{eq:EDV1}
\end{equation}
where the values of $E[L_{j}^{(1)}]$ and $E[g_{j}^{(1)}]$ have been
given by \cite{Wis2007}: 
\begin{equation}
E[L_{j}^{(1)}]=\left(e^{2\rho_{2}r_{I}}-1\right)\left(\frac{1}{\rho_{2}}-\frac{2r_{I}e^{-2\rho_{2}r_{I}}}{1-e^{-2\rho_{2}r_{I}}}\right),\label{eq:ELj}
\end{equation}
and 
\begin{equation}
E[g_{j}^{(1)}]=2r_{I}+\frac{1}{\rho_{2}}.\label{eq:Egj}
\end{equation}

As mentioned before, due to the memoryless property of exponential
distribution, $l_{0}$ has the same distribution as $l_{i}$ \cite{Feller},
i.e., we have: 
\begin{equation}
E[l_{0}]=\frac{1}{\rho_{2}}.\label{El0}
\end{equation}

Combining \eqref{eq:EDV1}-\eqref{El0}, we can obtain: 
\begin{equation}
E[D_{V1}]=\frac{\left[\frac{(d-2r_{I})(v_{1}+v_{2})}{v_{1}}+r_{0}-\frac{1}{\rho_{2}}\right]\left(1-e^{-2\rho_{2}r_{I}}\right)w_{I}}{v_{2}}\label{EDV1}
\end{equation}

By plugging equations \eqref{eq: DI} and \eqref{EDV1} into \eqref{eq: alternative expression of throughput},
we have the achievable throughput in Infrastructure-Limited regime,
denoted by $\eta_{1}$, as follows: 
\begin{equation}
\eta_{1}=\frac{2r_{I}w_{I}+c_{1}}{d},\label{eta1}
\end{equation}
where 
\[
c_{1}=\frac{\left[(d-2r_{I})(v_{1}+v_{2})+r_{0}v_{1}-\frac{v_{1}}{\rho_{2}}\right]\left(1-e^{-2\rho_{2}r_{I}}\right)w_{I}}{v_{2}}.
\]

\subsection{\label{subsec:V2V-Limited-Regime:}V2V-Limited Regime: $w_{I}\geq\frac{w_{V}v_{2}}{v_{1}+v_{2}}$}

Now we analyze the achievable throughput in the V2V-Limited regime.
Similar as subsection \ref{subsec:Infrastructure-Limited-Regime},
in this subsection, we first analyze the maximum amount of data that
can be received from \textit{helpers} by the VoI in one \textit{cycle}
by solving the optimization problem \eqref{eq:formulation of the optimization problem},
and then find the corresponding scheduling scheme to achieve this
maximum solution given fixed $n$ and $l_{i},i=1,...n-1$. Finally,
we extend to consider that $n$ and $l_{i},i=1,...n-1$ are random
values, corresponding to Poisson distribution of vehicles, and analyze
the achievable throughput under the proposed scheduling scheme.

\subsubsection{An analysis of the V2V communication process}

The following theorem summarizes the main result of this subsection.
\begin{thm}
\label{thm: throughput In-the-V2V-limited}In the V2V-Limited regime,
given fixed $n$ and $l_{i},i=1,...n-1$, the maximum amount of data
the VoI can receive from all $n$ helpers in one cycle is given by

\begin{equation}
\left(\sum_{i=1}^{n}Y_{i}\right)_{2}^{*}=\frac{\sum_{i=1}^{n-1}\min\{l_{i},2r_{0}\}+2r_{0}}{v_{1}+v_{2}}w_{V},\label{eq: optimal DV in V2V-Limited Regime}
\end{equation}
where $\left(\sum_{i=1}^{n}Y_{i}\right)_{2}^{*}$ is the respective
$\sum_{i=1}^{n}Y_{i}$ associated with its optimum value and we use
the subscript $2$ to mark the V2V-Limited regime.

Furthermore, there exists a V2I transmission scheme for helpers and
a V2V transmission scheme to reach the above maximum amount of received
data for the VoI, satisfying:
\begin{equation}
\begin{cases}
D_{2i}^{*}=\frac{\min\{l_{i},2r_{I}\}}{v_{2}}w_{I},i=1,2,...n-1\\
D_{2n}^{*}=\frac{2r_{I}}{v_{2}}w_{I}\\
Y_{2i}^{*}=\frac{\min\left\{ l_{i},2r_{0}\right\} }{v_{1}+v_{2}}w_{V},i=1,2...n-1\\
Y_{2n}^{*}=\frac{2r_{0}}{v_{1}+v_{2}}w_{V}
\end{cases}\label{eq: optimal scheme in V2V-Limited Regime}
\end{equation}
where $D_{2i}^{*}$ and $Y_{2i}^{*},i=1,...,n$ are the respective
$D_{i}$ and $Y_{i},i=1,...,n$ associated with the optimum solution.
\end{thm}
\begin{IEEEproof}
In the V2V-Limited regime, with conditions $w_{I}\geq\frac{w_{V}v_{2}}{v_{1}+v_{2}}$
and $r_{I}>r_{0}$, we have $\frac{2r_{0}}{v_{1}+v_{2}}w_{V}<\frac{2r_{I}}{v_{2}}w_{I}$
and $\frac{\min\{l_{i},2r_{0}\}}{v_{1}+v_{2}}w_{V}\leq\frac{\min\{l_{i},2r_{I}\}}{v_{2}}w_{I},i=1,...n-1$.
It follows that: 

\begin{equation}
\frac{\sum_{i=1}^{n-1}\min\{l_{i},2r_{0}\}+2r_{0}}{v_{1}+v_{2}}w_{V}<\frac{\sum_{i=1}^{n-1}\min\{l_{i},2r_{I}\}+2r_{I}}{v_{2}}w_{I},\label{ineq2 in V2V-Limited Regime}
\end{equation}
Then from constraints \eqref{eq:constraint on sum of Di}, \eqref{eq:Yi<Di},
\eqref{eq:constraint on sum of Yi} and inequality \eqref{ineq2 in V2V-Limited Regime},
we can conclude that the value of $\sum_{i=1}^{n}Y_{i}$ in optimization
problem \eqref{eq:formulation of the optimization problem} is upper
bound by $\frac{\sum_{i=1}^{n-1}\min\{l_{i},2r_{0}\}+2r_{0}}{v_{1}+v_{2}}w_{V}$.
In the following, we will show that this upper bound is exactly the
optimum solution of $\sum_{i=1}^{n}Y_{i}$ in the V2V-Limited regime
and can be reached under some scheduling scheme. 

Noting that $\frac{\sum_{i=1}^{n-1}\min\{l_{i},2r_{0}\}+2r_{0}}{v_{1}+v_{2}}w_{V}$
is the sum of $n$ separate components, with each component is not
larger than $\frac{2r_{0}}{v_{1}+v_{2}}w_{V}$. Therefore, when each
$Y_{i}$ is equal to the corresponding part of the $n$ component
forming the upper bound of $\sum_{i=1}^{n}Y_{i}$ , $\frac{\sum_{i=1}^{n-1}\min\{l_{i},2r_{0}\}+2r_{0}}{v_{1}+v_{2}}w_{V}$,
and each $D_{i}$ is equal to the corresponding part of the $n$ component
forming the upper bound of $\sum_{i=1}^{n}D_{i}$ shown in \eqref{eq:constraint on sum of Di},
i.e., when each $Y_{i}$ and $D_{i},i=1,...n$ are given by \eqref{eq: optimal scheme in V2V-Limited Regime},
the value of $\sum_{i=1}^{n}Y_{i}$ will reach its upper bound $\frac{\sum_{i=1}^{n-1}\min\{l_{i},2r_{0}\}+2r_{0}}{v_{1}+v_{2}}w_{V}$
while satisfying all constraints in optimization problem \eqref{eq:formulation of the optimization problem}.
This leads to the expression of \eqref{eq: optimal DV in V2V-Limited Regime}. 

\textcolor{black}{Now we show that there exists a scheduling scheme
to reach the maximum solution specified in \eqref{eq: optimal DV in V2V-Limited Regime}.
To this end, we show that \eqref{eq: optimal scheme in V2V-Limited Regime}
readily leads to the design of the scheduling scheme. Particularly,
the scheduling scheme schedules both }\textit{\textcolor{black}{helpers'}}\textcolor{black}{{}
V2I communication and V2V communications on a FIFO basis. Specifically,
the V2I transmission scheme for }\textit{\textcolor{black}{helpers}}\textcolor{black}{{}
is the same as that for the Infrastructure-Limited regime, which lead
to that each }\textit{\textcolor{black}{helper}}\textcolor{black}{{}
will receive an amount of data shown as each $D_{2i}^{*},i=1,2,...n$
in Eq. \eqref{eq: optimal scheme in V2V-Limited Regime}. For V2V
communications, the VoI starts to receive data from a }\textit{\textcolor{black}{helper}}\textcolor{black}{{}
once it enters this }\textit{\textcolor{black}{helper's}}\textcolor{black}{{}
coverage }\textcolor{black}{\emph{and}}\textcolor{black}{{} has retrieved
all data from the previous }\textit{\textcolor{black}{helper}}\textcolor{black}{{}
or has left the previous }\textit{\textcolor{black}{helper's}}\textcolor{black}{{}
coverage, and stops when the VoI leaves the coverage of the current
}\textit{\textcolor{black}{helper}}\textcolor{black}{{} or has retrieved
all data of the current }\textit{\textcolor{black}{helper}}\textcolor{black}{,
which lead to that the data the VoI receives from each }\textit{\textcolor{black}{helper}}\textcolor{black}{{}
is shown as each $Y_{2i}^{*},i=1,2,...n$ in Eq. \eqref{eq: optimal scheme in V2V-Limited Regime}.
Noting that Eq. \eqref{eq: optimal scheme in V2V-Limited Regime}
leads to the optimum solution \eqref{eq: optimal DV in V2V-Limited Regime},
therefore, it can be readily established that the aforementioned scheduling
scheme achieves the maximum amount of the received data for the VoI
specified in \eqref{eq: optimal DV in V2V-Limited Regime}.} This
completes the proof.
\end{IEEEproof}
Similarly as Theorem \ref{thm:In-the-infrastructure-limited-maximum-data},
Theorem \ref{thm: throughput In-the-V2V-limited} is also valid for
any value of $n$ and the corresponding $l_{i},i=1,...n-1$. 

\subsubsection{Throughput calculation}

On the basis of Theorem \ref{thm: throughput In-the-V2V-limited},
we now analyze the achievable throughput by the VoI considering that
both $n$ and the corresponding $l_{i},i=1,...n-1$ are random values.
In the V2V-Limited regime, we define a \emph{cluster} to be a maximal
set of \textit{helpers} located within road segment $\left[0,s\right]$
\emph{and} the distance between any two adjacent \textit{helpers}
is smaller than or equal to $2r_{0}$. The reason that we define the
\textit{clusters} differently from that in the Infrastructure-Limited
regime is that in this regime, it is the correlation in the V2V communication
process (and the associated difficulty in determining the joint distribution
of $\min\left\{ l_{i},2r_{0}\right\} ,i=1,...n-1$) that plays a dominating
effect on determining the achievable throughput. By defining \textit{clusters}
in the above way, within each \textit{cluster}, we have $\min\left\{ l_{i},2r_{0}\right\} =l_{i},i=1,2,...$.

In the V2V-Limited regime, the amount of data each \textit{cluster}
of \textit{helpers} delivers to the VoI, denoted by $R_{j}^{(2)}$,
can be calculated as follows: 
\begin{equation}
R_{j}^{(2)}=\frac{L_{j}^{(2)}+2r_{0}}{v_{1}+v_{2}}w_{V},j=1,...K_{2}\label{Ri3}
\end{equation}
where $L_{j}^{(2)},j=1,...K_{2}$ is the random variable representing
the length of the $j$-th \textit{cluster}, and $K_{2}$ is the random
integer representing the number of \textit{clusters} the VoI will
encounter in one \textit{cycle}. 

\textcolor{black}{Utilizing the same }\textit{\textcolor{black}{approximation}}\textcolor{black}{{}
method as that used to calculate $E[D_{V1}]$ in the Infrastructure-Limited
regime, i.e., }\textit{\textcolor{black}{approximately}}\textcolor{black}{{}
consider that $K_{2}$ and $L_{j}^{(2)}$ are independent in this
regime, the expected amount of data received by the VoI from }\textit{\textcolor{black}{helpers}}\textcolor{black}{{}
in one cycle in the V2V-Limited regime can be obtained as follows:}
\begin{equation}
E[D_{V2}]=\frac{s-E[l_{0}]}{E[L_{j}^{(2)}]+E[g_{j}^{(2)}]}\times\frac{E[L_{j}^{(2)}]+2r_{0}}{v_{1}+v_{2}}w_{V},\label{eq:EDV2}
\end{equation}
where $E[L_{i}^{(2)}]$ and $E[g_{i}^{(2)}]$ are given by: 
\begin{equation}
E[L_{j}^{(2)}]=\left(e^{2\rho_{2}r_{0}}-1\right)\left(\frac{1}{\rho_{2}}-\frac{2r_{0}e^{-2\rho_{2}r_{0}}}{1-e^{-2\rho_{2}r_{0}}}\right),\label{eq:ELj2}
\end{equation}

and 
\begin{equation}
E[g_{j}^{(2)}]=2r_{0}+\frac{1}{\rho_{2}}.\label{eq:Egj2}
\end{equation}

Combing \eqref{El0} and \eqref{eq:EDV2}-\eqref{eq:Egj2}, and simplifying
it, we have: 
\begin{equation}
E[D_{V2}]=\frac{\left[\frac{(d-2r_{I})(v_{1}+v_{2})}{v_{1}}+r_{0}-\frac{1}{\rho_{2}}\right]\left(1-e^{-2\rho_{2}r_{0}}\right)w_{V}}{v_{1}+v_{2}}.\label{EDV2}
\end{equation}

By plugging \eqref{eq: DI}, \eqref{EDV2} into \eqref{eq: alternative expression of throughput},
the achievable throughput in the V2V-Limited regime, denoted by $\eta_{2}$,
can be obtained as follows: 
\begin{equation}
\eta_{2}=\frac{2r_{I}w_{I}+c_{2}}{d},\label{eta2}
\end{equation}
where 
\[
c_{2}=\frac{\left[(d-2r_{I})(v_{1}+v_{2})+r_{0}v_{1}-\frac{v_{1}}{\rho_{2}}\right]\left(1-e^{-2\rho_{2}r_{0}}\right)w_{V}}{v_{1}+v_{2}}.
\]

\subsection{\label{subsec:Transitional-Regime:}Transitional Regime: $\frac{r_{0}w_{V}v_{2}}{r_{I}(v_{1}+v_{2})}<w_{I}<\frac{w_{V}v_{2}}{v_{1}+v_{2}}$}

Now we analyze the achievable throughput in the transitional regime
where the analysis is more intricate than that for the Infrastructure-Limited
and the V2V-Limited regime. Particularly, in the transitional regime,
both V2V communications and \textit{helpers'} V2I communications contribute
to determining the achievable throughput of the VoI. Therefore both
the correlation in the amount of data received by adjacent \textit{helpers}
from infrastructure and in the amount of data received by the VoI
from adjacent \textit{helpers}\textit{\emph{, as}} explained in Section
\ref{subsec:Problem-Formation}, need to be considered. This makes
finding the optimum solution for the optimization problem \eqref{eq:formulation of the optimization problem}
more challenging. Therefore, in this subsection, instead of analyzing
the exact achievable throughput, we analyze its upper and lower bound.
In the following, we will analyze the upper and the lower bound of
the achievable throughput separately. 

\subsubsection{\label{subsec:Upper-bound-of}Upper bound of the achievable throughput}

As shown in \eqref{eq:DV<min DV1 DV2}, an upper bound of $\sum_{i=1}^{n}Y_{i}$
is given by: 
\begin{equation}
\sum_{i=1}^{n}Y_{i}\leq\min\{D_{Vu1},D_{Vu2}\}\label{eq:upper bound of DV in transitional regime}
\end{equation}
That is, the upper bound of data amount received by the VoI from \textit{helpers
}is determined by the smaller value of the amount of data received
by the \textit{helpers}\emph{ }from their V2I communications, $D_{Vu1}$,
and the amount of data \textit{helpers} can deliver to the VoI in
V2V communications (without considering the limitation of the amount
of data they receive), $D_{Vu2}$. It is shown in Theorem \ref{thm:In-the-infrastructure-limited-maximum-data}
and Theorem \ref{thm: throughput In-the-V2V-limited} that $D_{Vu1}$
and $D_{Vu2}$ are exactly the maximum amount of data the VoI can
receive from \textit{helpers} in the Infrastructure-Limited regime
and V2V-Limited regime respectively. Therefore, according to the throughput
calculation analysis given in subsection \ref{subsec:Infrastructure-Limited-Regime:-}
and \ref{subsec:V2V-Limited-Regime:}, when $\sum_{i=1}^{n}Y_{i}$
is upper bounded by $D_{Vu1}$ (or $D_{Vu2})$, the corresponding
achievable throughput of the VoI will be upper bounded by the achievable
throughput in the Infrastructure-Limited regime, $\eta_{1}$, (or
the achievable throughput in the V2V-Limited regime,$\eta_{2}$,).
It follows that an upper bound of the achievable throughput by the
VoI in the transitional regime, denoted by $\eta_{3u}$, is given
by: 
\begin{align}
\eta_{3u} & =\min\{\eta_{1},\eta_{2}\}=\begin{cases}
\eta_{1}, & \eta_{1}\leq\eta_{2}\\
\eta_{2}, & \eta_{1}>\eta_{2}
\end{cases}\label{eq:eta3u-1}
\end{align}
Putting \eqref{eta1} and \eqref{eta2} into \eqref{eq:eta3u-1} and
simplifying it, we have: 
\begin{align}
\eta_{3u} & =\begin{cases}
\frac{2r_{I}w_{I}+c_{1}}{d},\;\;\frac{r_{0}w_{V}v_{2}}{r_{I}(v_{1}+v_{2})}<w_{I}\leq\frac{1-e^{-2\rho_{2}r_{0}}}{1-e^{-2\rho_{2}r_{I}}}\times\frac{w_{V}v_{2}}{v_{1}+v_{2}}\\
\frac{2r_{I}w_{I}+c_{2}}{d},\;\;\frac{1-e^{-2\rho_{2}r_{0}}}{1-e^{-2\rho_{2}r_{I}}}\times\frac{w_{V}v_{2}}{v_{1}+v_{2}}<w_{I}<\frac{w_{V}v_{2}}{v_{1}+v_{2}}
\end{cases}\label{eta3u}
\end{align}
with $c_{1}$ and $c_{2}$ have been given in the earlier analysis.
\begin{rem}
\label{Remark for lower bound}Equation \eqref{eta3u} shows that
$\frac{1-e^{-2\rho_{2}r_{0}}}{1-e^{-2\rho_{2}r_{I}}}\times\frac{w_{V}v_{2}}{v_{1}+v_{2}}$
is a transition point for the value of $w_{I}$ to determine the upper
bound of achievable throughput in the transitional regime, whose value
depends on the \textit{helpers'} density $\rho_{2}$. Specifically,
when $\rho_{2}\rightarrow0$, $\frac{1-e^{-2\rho_{2}r_{0}}}{1-e^{-2\rho_{2}r_{I}}}\times\frac{w_{V}v_{2}}{v_{1}+v_{2}}\rightarrow\frac{r_{0}w_{V}v_{2}}{r_{I}(v_{1}+v_{2})}$;
when $\rho_{2}$ increases, the gap between $\frac{1-e^{-2\rho_{2}r_{0}}}{1-e^{-2\rho_{2}r_{I}}}\times\frac{w_{V}v_{2}}{v_{1}+v_{2}}$
and $\frac{r_{0}w_{V}v_{2}}{r_{I}(v_{1}+v_{2})}$ becomes larger and
the gap between $\frac{1-e^{-2\rho_{2}r_{0}}}{1-e^{-2\rho_{2}r_{I}}}\times\frac{w_{V}v_{2}}{v_{1}+v_{2}}$
and $\frac{w_{V}v_{2}}{v_{1}+v_{2}}$ becomes smaller; and when $\rho_{2}\rightarrow\infty$,
$\frac{1-e^{-2\rho_{2}r_{0}}}{1-e^{-2\rho_{2}r_{I}}}\times\frac{w_{V}v_{2}}{v_{1}+v_{2}}\rightarrow\frac{w_{V}v_{2}}{v_{1}+v_{2}}$. 
\end{rem}

\subsubsection{\label{subsec:Lower-bound-of}Lower bound of the achievable throughput}

In this subsection, we first analyze the lower bound of the maximum
amount of data that can be received from \textit{helpers} by the VoI
and the corresponding scheduling scheme to achieve this lower bound
given fixed $n$ and $l_{i},i=1,...n-1$. Then we extend to consider
that $n$ and $l_{i},i=1,...n-1$ are random values, corresponding
to Poisson distribution of vehicles, and analyze the lower bound of
the achievable throughput. 
\begin{thm}
\label{thm: lower bound of throughput in transitional regime}In the
transitional regime, given fixed $n$ and $l_{i},i=1,...n-1$, a lower
bound of the maximum amount of data the VoI can receive from $n$
helpers in one cycle is given by 
\begin{equation}
\left(\sum_{i=1}^{n}Y_{i}\right)_{3}^{*}\geq\sum_{i=1}^{n-1}\min\left\{ \frac{l_{i}}{v_{2}}w_{I},\frac{2r_{0}}{v_{1}+v_{2}}w_{V}\right\} +\frac{2r_{0}}{v_{1}+v_{2}}w_{V}\label{eq:lower bound of DV in transitional regime}
\end{equation}
where $\left(\sum_{i=1}^{n}Y_{i}\right)_{3}^{*}$ is the respective
$\sum_{i=1}^{n}Y_{i}$ associated with its optimum value and we use
the subscript $3$ to mark the transitional regime.

Furthermore, there exists a V2I transmission scheme for helpers and
a V2V transmission scheme to achieve the above lower bound of the
maximum amount of data for the VoI, satisfying: 
\begin{equation}
\begin{cases}
D_{3i}^{*}=\frac{\min\{l_{i},2r_{I}\}}{v_{2}}w_{I}, & i=1,2,...n-1\\
D_{3n}^{*}=\frac{2r_{I}}{v_{2}}w_{I}\\
Y_{3i}^{*}=\min\left\{ \frac{l_{i}w_{I}}{v_{2}},\frac{2r_{0}w_{V}}{v_{1}+v_{2}}\right\} , & i=1,2...n-1\\
Y_{3n}^{*}=\frac{2r_{0}}{v_{1}+v_{2}}w_{V}
\end{cases}\label{eq:optimal transmitting scheme for transitional scheme}
\end{equation}
where $D_{3i}^{*}$ and $Y_{3i}^{*},i=1,...,n$ are the respective
$D_{i}$ and $Y_{i},i=1,...,n$ associated with the optimum solution.
\end{thm}
\begin{IEEEproof}
We find the lower bound of the maximum amount of data received by
the VoI from \textit{$n$ helpers} in one \textit{cycle} by constructing
a specific V2I transmission scheme and analyze the corresponding value
of $\sum_{i=1}^{n}Y_{i}$ achieved under this scheme. As this value
of $\sum_{i=1}^{n}Y_{i}$ is obtained under a specific V2I transmission
scheme, it may not be the maximum value for the original optimization
problem \eqref{eq:formulation of the optimization problem} because
of a lack of consideration of all possible V2I transmission schemes
for \textit{helpers}, but will form a lower bound of the maximum value
of $\sum_{i=1}^{n}Y_{i}$ for the the original optimization problem
\eqref{eq:formulation of the optimization problem}. In the following,
we will first construct a specific V2I transmission scheme, and then
analyze the optimum amount of data received by the VoI from \textit{helpers}
under this specific V2I transmission scheme, as well as finding a
corresponding V2V transmission scheme to reach the lower bound specified
in the theorem. 

It has been described in the proof of Theorem \ref{thm:In-the-infrastructure-limited-maximum-data}
and Theorem \ref{thm: throughput In-the-V2V-limited} that the V2I
transmission scheme for \textit{helpers} to reach the corresponding
optimum throughput under the Infrastructure-Limited and the V2V-Limited
regime are the same, and this V2I transmission scheme satisfies the
following equations: 
\begin{equation}
\begin{cases}
D_{i}=\frac{\min\{l_{i},2r_{I}\}}{v_{2}}w_{I}, & i=1,2,...n-1\\
D_{n}=\frac{2r_{I}}{v_{2}}w_{I}.
\end{cases}\label{eq: V2I's transmitting scheme}
\end{equation}
We adopt this same V2I transmission scheme for \textit{helpers} in
the transitional regime as well. It follows that the amount of data
received by each \textit{helper} from infrastructure, $D_{i}$, $i=1,...,n$,
is given by \eqref{eq: V2I's transmitting scheme}. With condition
$\frac{r_{0}w_{V}v_{2}}{r_{I}(v_{1}+v_{2})}<w_{I}<\frac{w_{V}v_{2}}{v_{1}+v_{2}}$
for the transitional regime, constraints \eqref{eq:constraint on Yi}
and \eqref{eq:Yi<Di} in the optimization problem \eqref{eq:formulation of the optimization problem}
can be replaced with a tighter constraint after putting in \eqref{eq: V2I's transmitting scheme},
which are shown as follows:
\begin{alignat}{1}
Y_{i} & \leq\min\left\{ D_{i},\frac{2r_{0}}{v_{1}+v_{2}}w_{V}\right\} \nonumber \\
 & =\min\left\{ \frac{l_{i}}{v_{2}}w_{I},\frac{2r_{I}}{v_{2}}w_{I},\frac{2r_{0}}{v_{1}+v_{2}}w_{V}\right\} \nonumber \\
 & =\min\left\{ \frac{l_{i}}{v_{2}}w_{I},\frac{2r_{0}}{v_{1}+v_{2}}w_{V}\right\} ,i=1,2,...n-1\label{tight bound of Yi in transitional regime}
\end{alignat}

and 
\begin{equation}
0\leq Y_{n}\leq\frac{2r_{0}}{v_{1}+v_{2}}w_{V}\label{tight bound of Yn in transitional regime}
\end{equation}

From \eqref{tight bound of Yi in transitional regime} and \eqref{tight bound of Yn in transitional regime},
the upper bound of $\sum_{i=1}^{n}Y_{i}$ is given by $\sum_{i=1}^{n-1}\min\left\{ \frac{l_{i}}{v_{2}}w_{I},\frac{2r_{0}}{v_{1}+v_{2}}w_{V}\right\} +\frac{2r_{0}}{v_{1}+v_{2}}w_{V}$.
In the following, we will show that this upper bound is exactly the
optimum solution of $\sum_{i=1}^{n}Y_{i}$ under the adopted V2I transmission
scheme and we can find a corresponding V2V transmission scheme to
achieve this upper bound. 

With condition $w_{I}<\frac{w_{V}v_{2}}{v_{1}+v_{2}}$, we have:
\begin{equation}
\min\left\{ \frac{l_{i}}{v_{2}}w_{I},\frac{2r_{0}}{v_{1}+v_{2}}w_{V}\right\} \leq\min\left\{ \frac{l_{i}}{v_{1}+v_{2}}w_{V},\frac{2r_{0}}{v_{1}+v_{2}}w_{V}\right\} 
\end{equation}
This follows that: 
\begin{alignat}{1}
\sum_{i=1}^{n}Y_{i} & \leq\sum_{i=1}^{n-1}\min\left\{ \frac{l_{i}}{v_{2}}w_{I},\frac{2r_{0}}{v_{1}+v_{2}}w_{V}\right\} +\frac{2r_{0}}{v_{1}+v_{2}}w_{V}\nonumber \\
 & \leq\frac{\sum_{i=1}^{n-1}\min\left\{ l_{i},2r_{0}\right\} +2r_{0}}{v_{1}+v_{2}}w_{V},
\end{alignat}
which shows that when $\sum_{i=1}^{n}Y_{i}$ is not larger than $\sum_{i=1}^{n-1}\min\left\{ \frac{l_{i}}{v_{2}}w_{I},\frac{2r_{0}}{v_{1}+v_{2}}w_{V}\right\} +\frac{2r_{0}}{v_{1}+v_{2}}w_{V}$,
the constraint \eqref{eq:constraint on sum of Yi} in optimization
problem \eqref{eq:formulation of the optimization problem} will also
be satisfied. Thus, when each $Y_{i},i=1,...n$, is equal to its upper
bound shown in \eqref{tight bound of Yi in transitional regime} and
\eqref{tight bound of Yn in transitional regime}, and when each $D_{i},i=1,..n$
is given by \eqref{eq: V2I's transmitting scheme}, i.e., when each
$Y_{i}$ and $D_{i},i=1,...n$ are given by \eqref{eq:optimal transmitting scheme for transitional scheme},
the value of $\sum_{i=1}^{n}Y_{i}$ will reach its upper bound $\sum_{i=1}^{n-1}\min\left\{ \frac{l_{i}}{v_{2}}w_{I},\frac{2r_{0}}{v_{1}+v_{2}}w_{V}\right\} +\frac{2r_{0}}{v_{1}+v_{2}}w_{V}$
while satisfying all constraints in the optimization problem \eqref{eq:formulation of the optimization problem}.
This leads to \eqref{eq:lower bound of DV in transitional regime}.

Now we show that there exists a V2V transmission scheme to achieve
the lower bound specified in \eqref{eq:lower bound of DV in transitional regime}.
To this end, it can be readily shown from \eqref{eq:optimal transmitting scheme for transitional scheme}
that the same V2V transmission scheme described in the proof of Theorem
\ref{thm:In-the-infrastructure-limited-maximum-data} satisfies \eqref{eq:optimal transmitting scheme for transitional scheme},
therefore can realize the maximum amount of the received data for
the VoI specified in \eqref{eq:lower bound of DV in transitional regime}
under the specific V2I transmission scheme. This completes the proof.
\end{IEEEproof}
On the basis of Theorem \ref{thm: lower bound of throughput in transitional regime},
we now analyze the lower bound of the achievable throughput by the
VoI considering that both $n$ and the corresponding $l_{i},i=1,...n-1$
are random values. Similarly to the analysis in subsections \ref{subsec:Infrastructure-Limited-Regime:-}
and \ref{subsec:V2V-Limited-Regime:}, calculating the lower bound
of the achievable throughput directly according to \eqref{eq:lower bound of DV in transitional regime}
is challenging due to complexity associated with analyzing the joint
distribution of $\min\left\{ \frac{l_{i}}{v_{2}}w_{I},\frac{2r_{0}}{v_{1}+v_{2}}w_{V}\right\} ,i=1,...n-1$.
In this regime, we define a \textit{cluster} to be a maximal set of
\textit{helpers}  located within road segment $\left[0,s\right]$
and the distance between any two adjacent \textit{helpers} is smaller
than or equal to $\frac{2r_{0}w_{V}v_{2}}{w_{I}(v_{1}+v_{2})}$. It
follows that within each \textit{cluster}, $\min\left\{ \frac{l_{i}}{v_{2}}w_{I},\frac{2r_{0}}{v_{1}+v_{2}}w_{V}\right\} =\frac{l_{i}}{v_{2}}w_{I},i=1,2,...$,
therefore removing the above challenge. 

\textcolor{black}{Utilizing the same }\textit{\textcolor{black}{approximation}}\textcolor{black}{{}
method as that used to calculate $E[D_{V1}]$ and $E[D_{V2}]$ in
the Infrastructure-Limited regime and the V2V-Limited regime respectively,
the lower bound of the achievable throughput in transitional regime,
denoted by $\eta_{3l}$, is obtained as follows:}

\begin{equation}
\eta_{3l}=\frac{2r_{I}w_{I}+c_{3}}{d},\label{lower bound of throughput in transitional regime}
\end{equation}
where $c_{3}=\frac{\left[(d-2r_{I})(v_{1}+v_{2})+r_{0}v_{1}-\frac{v_{1}}{\rho_{2}}\right]\left(1-e^{-\frac{2\rho_{2}r_{0}w_{V}v_{2}}{w_{I}(v_{1}+v_{2})}}\right)w_{I}}{v_{2}}$.

\section{Simulation and Discussion\label{sec:Simulation-and-Discussion}}

In this section we use simulations conducted in Matlab to verify the
accuracy of the analysis and establish the applicability of the theoretical
analysis for more general scenarios beyond the ideal assumptions (e.g.,
constant speed, unit disk model, constant channel condition, .etc)
used in the analysis. Specifically, 20 infrastructure points are regularly
deployed along the highway and the distance between adjacent infrastructure
point, $d$, is varied from 2km to 50km. The \textit{helpers}\emph{'}
density $\rho_{2}$ varies from 0 to 0.1veh/m and the speed of the
VoI and \textit{helpers} are $v_{1}$=15m/s and $v_{2}$=25m/s respectively.
The radio range of infrastructure and vehicles are 500m and 250m (typical
radio ranges using DSRC \cite{Haibo14}) respectively. The transmission
rate of V2V communications is $w_{V}$=5Mb/s and the transmission
rate of V2I communications $w_{I}$ varies from 0 to 10Mb/s to allow
us to cover all three regimes. Each simulation is repeated 2000 times
and the average value is shown in the plot.

\begin{figure}
\centering{\subfigure[Infrastructure-Limited Regime]{\label{SIMU-ANAL-Infrastructure-Limited Regime}\includegraphics[width=8cm]{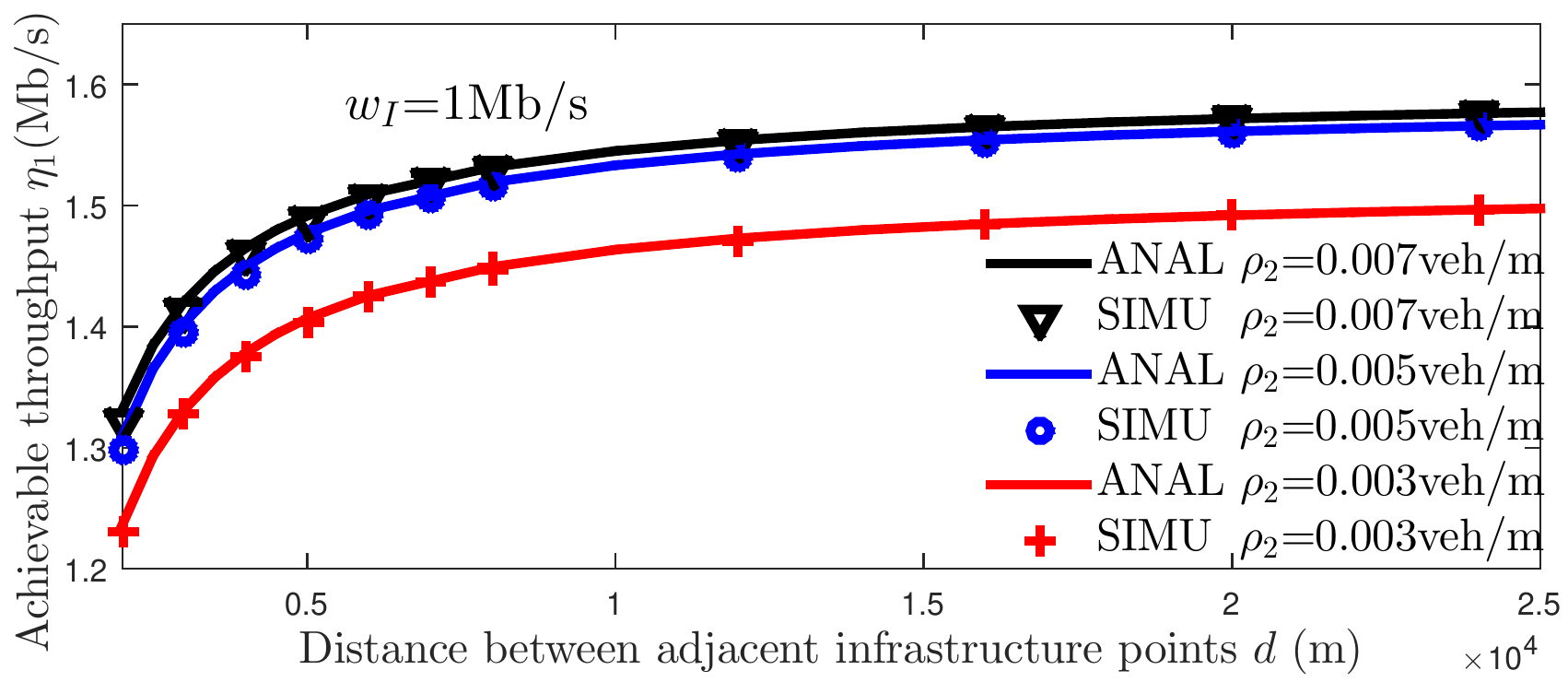}}\\
\subfigure[V2V-Limited Regime]{\label{SIMU-ANAL-V2V-Limitel Regime}\includegraphics[width=8.1cm]{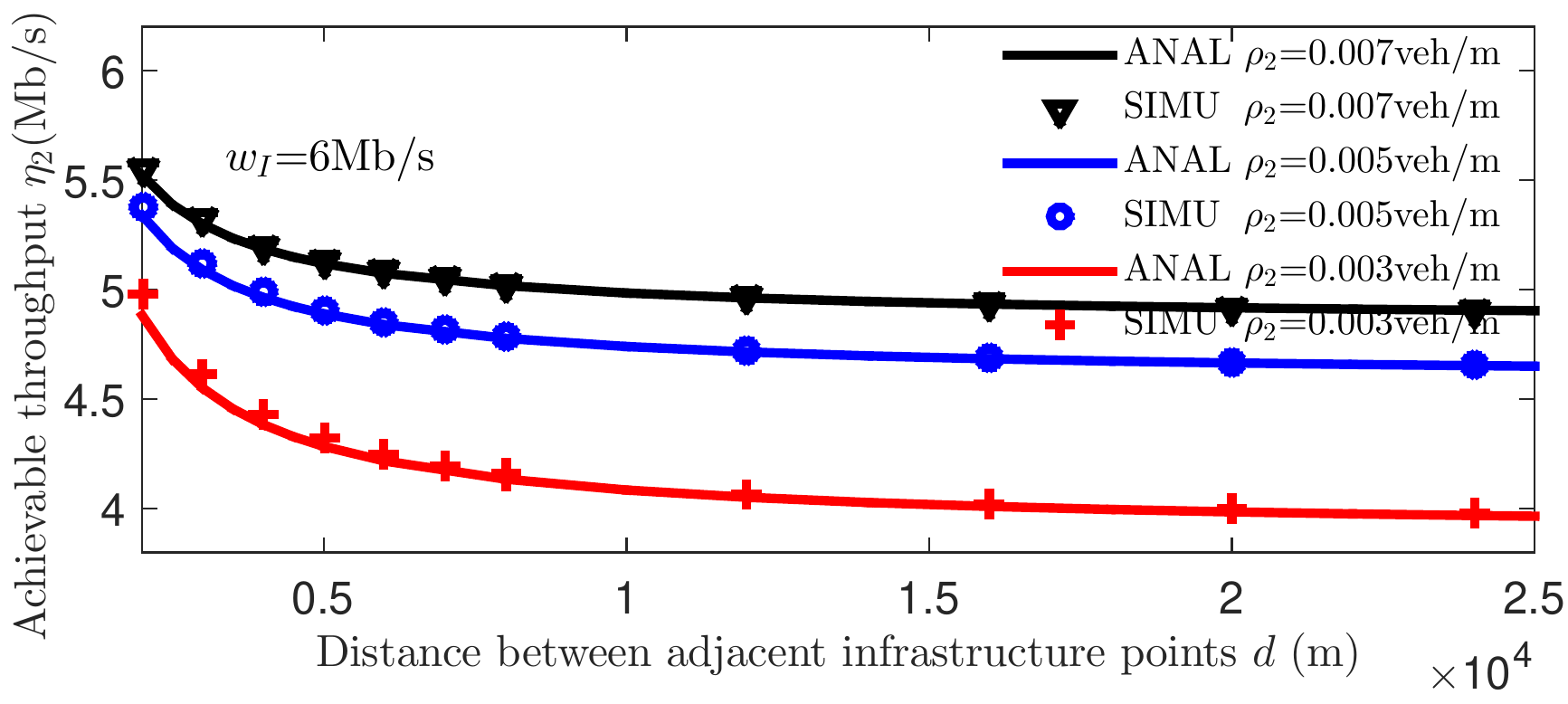}}

\subfigure[Transitional Regime]{\label{SIMU-ANAL-Transitional  Regime}\includegraphics[width=8cm]{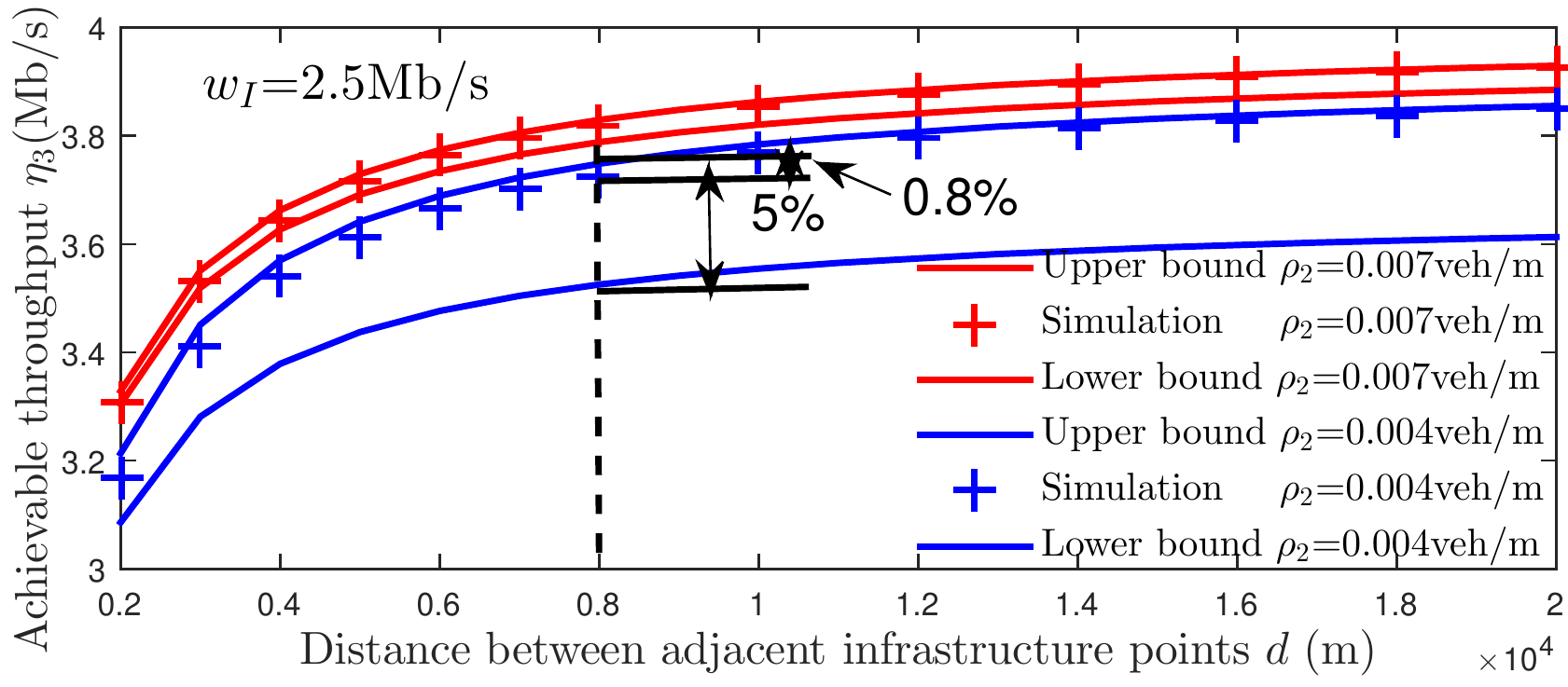}}\\
}

\caption{A comparison between our analytical results and the simulation results
under each regime, with different \textit{helpers'} density $\rho_{2}$. }

\label{SIMU-ANAL}
\end{figure}

Fig. \ref{SIMU-ANAL} shows a comparison between analytical results
and simulation results under each regime. Specifically, Fig. \ref{SIMU-ANAL-Infrastructure-Limited Regime}
and Fig. \ref{SIMU-ANAL-V2V-Limitel Regime} compare the achievable
throughput obtained from our analysis and the simulation result, where
the simulation is conducted assuming the scheduling scheme described
in the proof of Theorem \ref{thm:In-the-infrastructure-limited-maximum-data}
and Theorem \ref{thm: throughput In-the-V2V-limited} respectively.
Fig. \ref{SIMU-ANAL-Transitional  Regime} compares the upper and
lower bound of the achievable throughput obtained from analysis and
the optimum throughput in the simulation. It is shown that in the
Infrastructure-Limited and V2V-Limited regime, the analytical results
match very well with simulations especially when the distance of two
neighboring infrastructure, $d$, is large. This confirms that the
approximations used in the earlier analysis to obtain the analytical
results have negligible impact on the accuracy of the analytical results.
In the transitional regime, there is a small gap between the simulated
optimum throughput and its upper and lower bound we obtained, e.g,
when $\rho_{2}$=0.004veh/m and $d$=8km in this case, the difference
between the optimum throughput from the simulation and its upper (or
lower) bound is only around 0.8\% (or 5\%), and the gap decreases
with the increase of \textit{helpers'} density. This shows that even
though the upper bound in the transitional regime is not achievable,
it is quite close to the optimum throughput. 

Interestingly, Fig. \ref{SIMU-ANAL-Infrastructure-Limited Regime}
and \ref{SIMU-ANAL-Transitional  Regime} show that in the Infrastructure-Limited
regime and the transitional regime, the achievable throughput increases
when $d$ increases while Fig. \ref{SIMU-ANAL-V2V-Limitel Regime}
shows that in the V2V-Limited regime, the achievable throughput decreases
when $d$ increases. This can be explained that while keeping other
parameters constant, an increase in $d$ on one hand will improve
the amount of data received by the VoI from V2V communications, which
improves the achievable throughput; on the other hand, it will increase
the amount of time spent in one \textit{cycle} by the VoI, which reduces
the achievable throughput. When $w_{I}$ is small, the amount of data
received by the VoI from V2V communications is small due to the limitation
of the amount of data received by \textit{helpers} from infrastructure.
Therefore, an increase in $d$ will lead to an larger rate of increase
in the total amount of received data by the VoI than the rate of increase
in the amount of time spent in one \textit{cycle}, which results in
the overall increase of the achievable throughput, shown as Fig. \ref{SIMU-ANAL-Infrastructure-Limited Regime}
and \ref{SIMU-ANAL-Transitional  Regime}. However, when $w_{I}$
is large, the amount of data the VoI can receive from V2V communications
is comparatively large, an increase in $d$ has marginal impact on
the data amount received by the VoI. Therefore, an increase in $d$
will lead to an smaller rate of increase in the total amount of received
data by the VoI than the rate of increase in the amount of time spent
in one \textit{cycle}, which results in the overall decrease of the
achievable throughput, shown as Fig. \ref{SIMU-ANAL-V2V-Limitel Regime}.

Fig. \ref{SIMU-ANAL} also gives insight into the optimum choice of
distance between infrastructure points. It is obvious from these figures
that when $d$ increases beyond a certain threshold, e.g., $d$=10km
in our case, an increase in $d$ has limited impact on the achievable
throughput. This can be explained by the fact that when $d$ is small
($d<$10km in the simulation), the amount of data received by the
VoI from V2V communications is relatively small compared with that
received from V2I communications, especially when traffic density
is low (here average $\rho_{2}$=0.005veh/m). It follows that the
VoI's achievable throughput is mainly dominated by its V2I communications.
However, with the increase of $d$, the increase of data received
from V2V communications makes V2I communication's dominating impact
subdued, which in turn leads to the subtle variation in the throughput.

\begin{figure}
\centering \includegraphics[width=8cm]{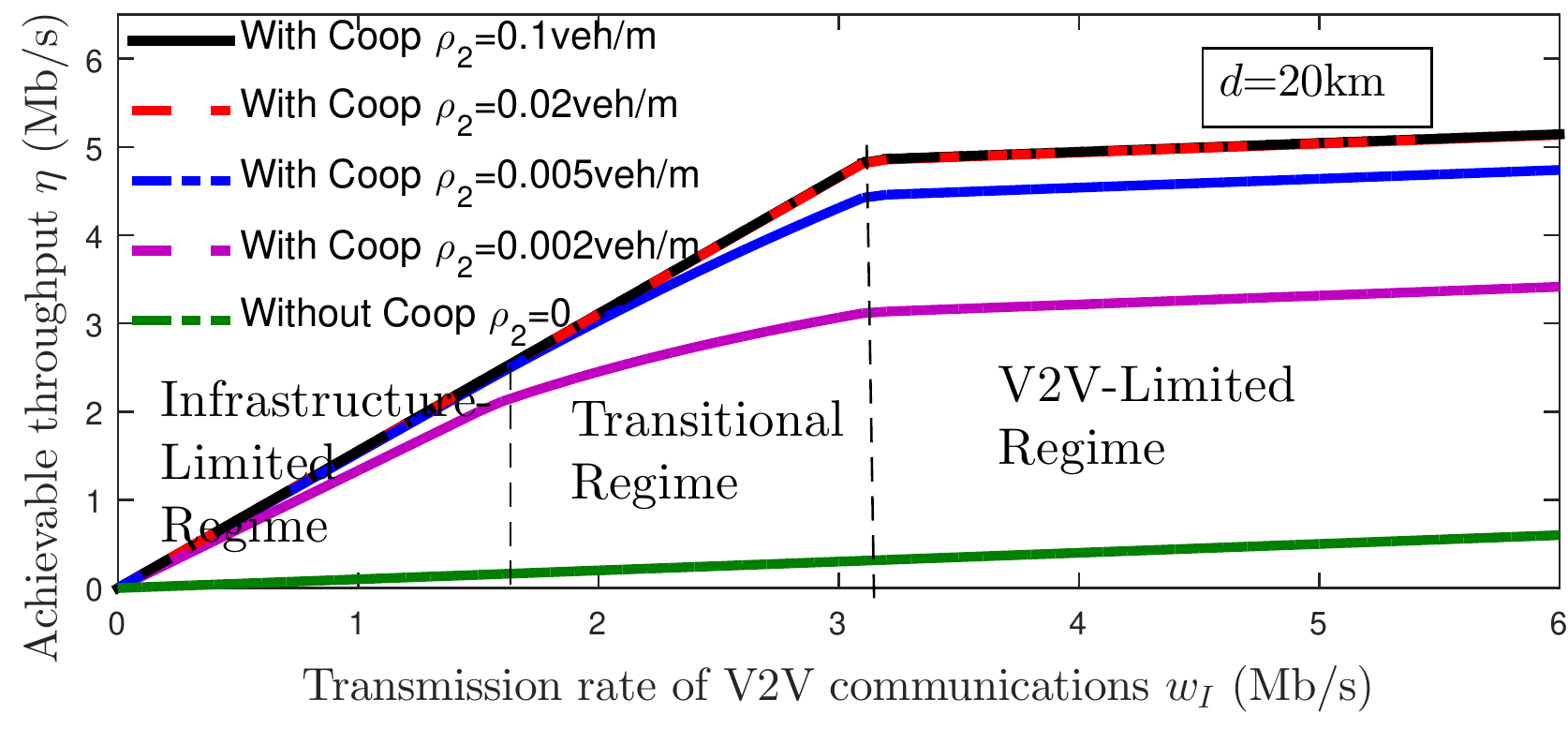} \caption{A comparison between the throughput achieved from vehicular networks
with and without cooperative communication by setting \textit{helpers}'
density $\rho_{2}$ as 0.1veh/m (near-capacity), 0.02veh/m (congested),
0.005veh/m (low density), 0.002veh/m (low density) and 0 (without
cooperation) respectively. }
\label{rate_wi_rho} 
\end{figure}

Fig. \ref{rate_wi_rho} compares the achievable throughput (the lower
bound for the transitional-regime is used) using our cooperative communication
strategy (labeled as With coop) with its non-cooperative counterpart
(labeled as Without coop). The non-cooperative counterpart is conducted
by setting the \textit{helpers'} density $\rho_{2}=0$ because when
there is no \textit{helpers} in the vehicular network, there will
be no cooperative communications. It is shown that even when \textit{helpers'}
density is low, e.g., $\rho_{2}$=0.002veh/m, the throughput achieved
by utilizing our cooperative communication strategy is around 15 times
larger when $w_{I}$=3Mb/s and around 10 times larger when $w_{I}$=6Mb/s
than that achieved without cooperative communications. This gives
an important conclusion that our cooperative communication strategy
can significantly improve the throughput even when\emph{ }vehicular
density is rather low. 

Fig. \ref{rate_wi_rho} also reveals the relationship between the
achievable throughput and \textit{helpers'} density $\rho_{2}$. Importantly,
we can see that a higher density is beneficial to the throughput because
a higher $\rho_{2}$ will enhance the connectivity of vehicular networks,
which leads to higher chance of V2V cooperative communications. However,
when $\rho_{2}$ increases beyond a certain threshold, e.g., $\rho_{2}$=0.005veh/m
in this case, a further increase in $\rho_{2}$ has only marginal
impact on the achievable throughput. This is due to the fact that
when $\rho_{2}$ is large enough for the VoI to find at least one
\textit{helper} in its coverage at any time point, increasing the
density (which will lead to more\emph{ }\textit{helpers }within the
VoI's coverage at one time) is no longer helpful to improve the throughput
because the VoI can only receive data from one vehicle at one time
and the total amount of time the VoI can receive data from V2V communication
will be the same.

\begin{figure}
\centering \includegraphics[width=8.1cm]{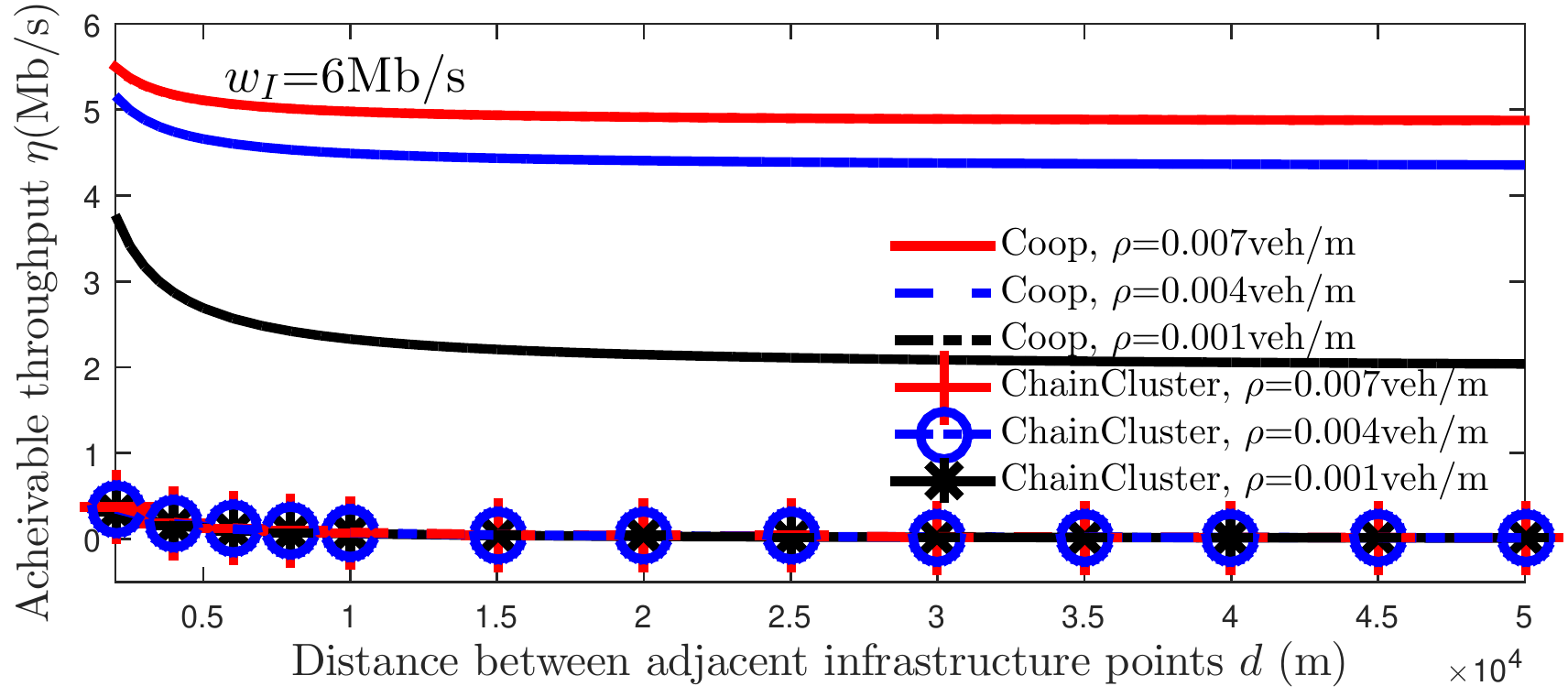} \caption{A comparison between the throughput achieved from our proposed strategy
and that from the strategy proposed in \cite{Haibo14}. }
\label{Comp_Cluster} 
\end{figure}

\begin{figure}
\centering \includegraphics[width=8cm]{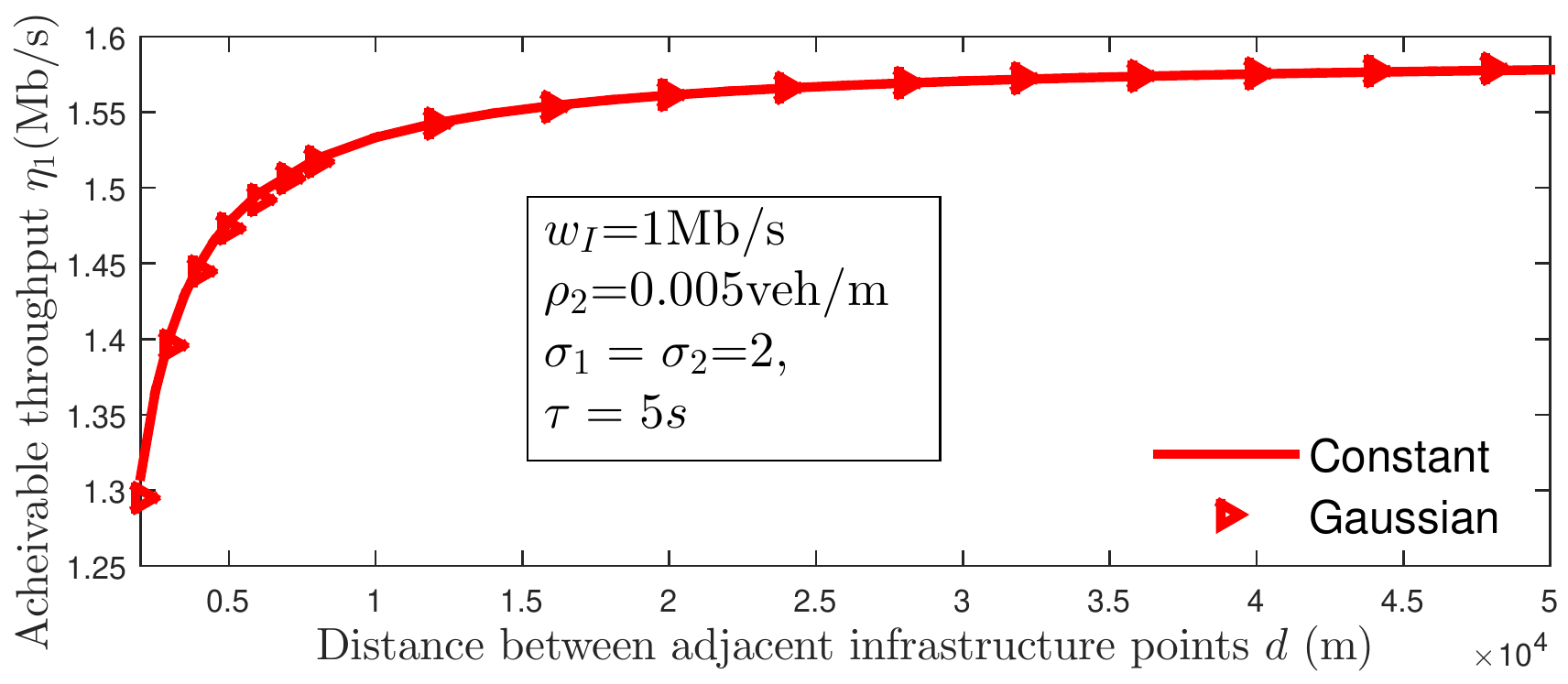} \caption{A comparison between throughput achieved from the constant speed model
and the time-varying speed model which follows Gaussian distribution
with mean value $v_{1}=15$m/s and $v_{2}=25$m/s, variance $\sigma_{1}=\sigma_{2}=2$,
and the speed-change time interval $\tau$=5s.}
\label{vary_speed} 
\end{figure}

\textcolor{black}{Fig. \ref{Comp_Cluster} compares the achievable
throughput assuming our proposed cooperative communication strategy
(labeled as Coop) with that assuming the cooperative strategy proposed
in \cite{Haibo14} (labeled as ChainCluster) in the V2V-Limited regime.
Specifically, the strategy proposed in \cite{Haibo14} utilized vehicles
moving in the same direction as the target vehicle (VoI) to form a
cluster to help the VoI's download. A vehicle can be chosen into the
cluster if and only if it can connect to the VoI via a multi-hop path.
It can be seen that the throughput achieved by the VoI assuming our
cooperative communication strategy is much larger than that achieved
assuming the strategy proposed in \cite{Haibo14}. This is due to
the fact that in \cite{Haibo14}, the authors only used the cooperation
among vehicles moving in the same direction and within the same cluster
of the VoI, while in our strategy, both cooperation among infrastructure
and cooperation of all vehicles traveling in the opposite direction
of the VoI are fully utilized to help the VoI's download, which significantly
improves the achievable throughput of the VoI.}

Fig. \ref{vary_speed} shows a comparison of the achievable throughput
from the constant speed model (labeled as Constant Speed) and the
time-varying speed model (labeled as Gaussian Speed) under the Infrastructure-Limited
regime. The time-varying speed of vehicles in each lane follows Gaussian
distributions, defined as: $v_{1}^{'}\sim N(v_{1},\sigma_{1}^{2})$
and $v_{2}^{'}\sim N(v_{2},\sigma_{2}^{2})$, where $v_{1}$ and $v_{2}$
are the constant speed we used in our analysis, and $\sigma_{1}^{2}$
and $\sigma_{2}^{2}$ are the mean variance of the mean speed $v_{1}$
and $v_{2}$ respectively. To model the slight deviations from the
mean speed, we set $\sigma_{1}$=$\sigma_{2}$=2 and the speed-change
time interval $\tau$=5s. The figure shows that when individual vehicular
speed deviates slightly from the mean speed, it has marginal impact
on the achievable throughput. 

\begin{figure}
\centering \includegraphics[width=8cm]{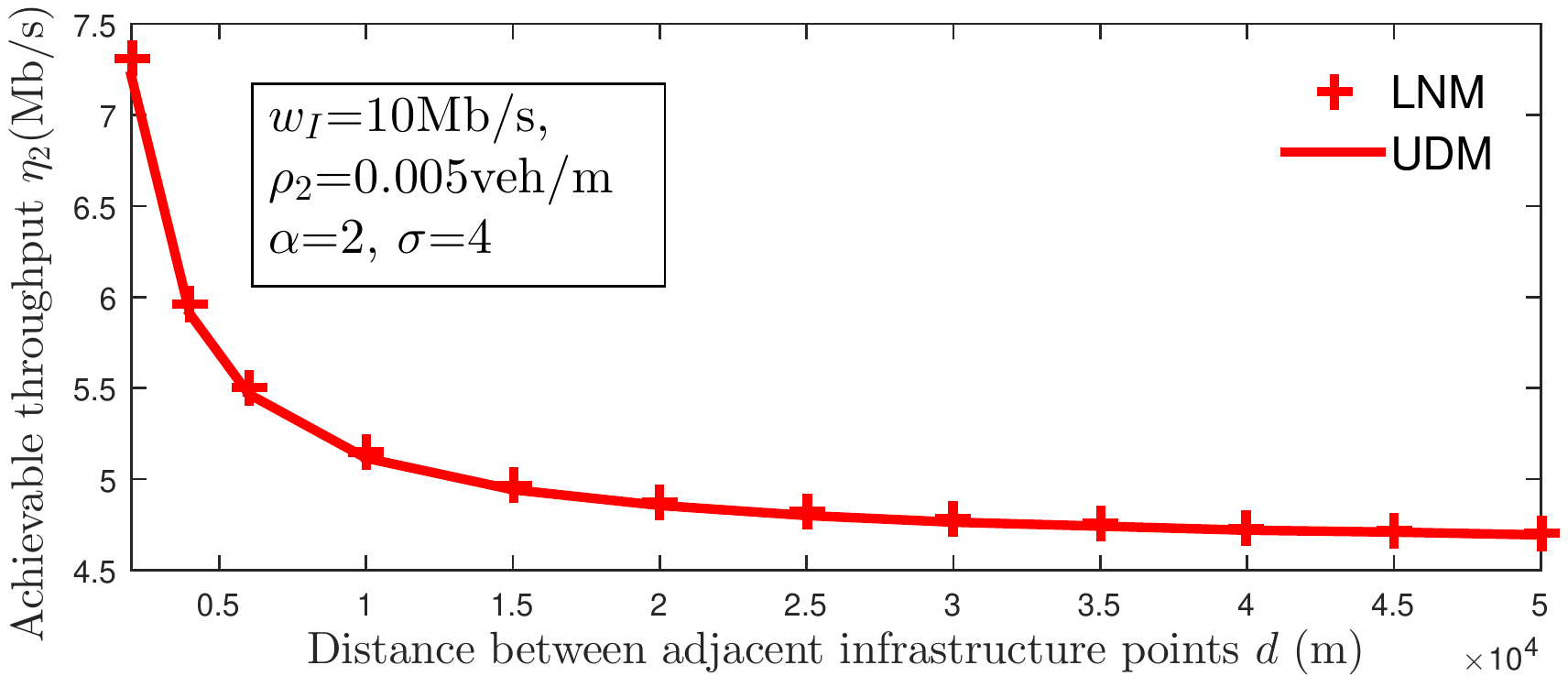} \caption{A comparison between throughput achieved from the unit disk model
and the log-normal connection model with path loss exponent $\alpha$=2
and standard variance $\sigma$=4.}
\label{LSM_UDM} 
\end{figure}

\textcolor{black}{Fig. \ref{LSM_UDM} gives a comparison of throughput
achieved assuming the unit disk model (labeled as UDM) and that assuming
the log-normal connection model (labeled as LNM) in the V2V-Limited
regime, and shows that the unit disk model assumption has little impact
on the throughput. The parameters of log-normal connection model are
set as: path loss exponent $\alpha$=2 and standard deviation $\sigma$=4
\cite{ZijieICC14}. It is shown that the system assuming the log-normal
connection model has a slightly higher achievable throughput than
that assuming the unit disk model, which coincides with the results
in our previous paper \cite{ZijieICC14} that log-normal connection
model is beneficial to information delivery in vehicular networks.
The reason behind this phenomenon is that the log-normal connection
model introduces a Gaussian variation of the transmission range around
the mean value, which implies a higher chance for the VoI to be connected
to }\textit{\textcolor{black}{helpers}}\textcolor{black}{{} separated
further away.}

\begin{figure}
\centering \includegraphics[width=8cm]{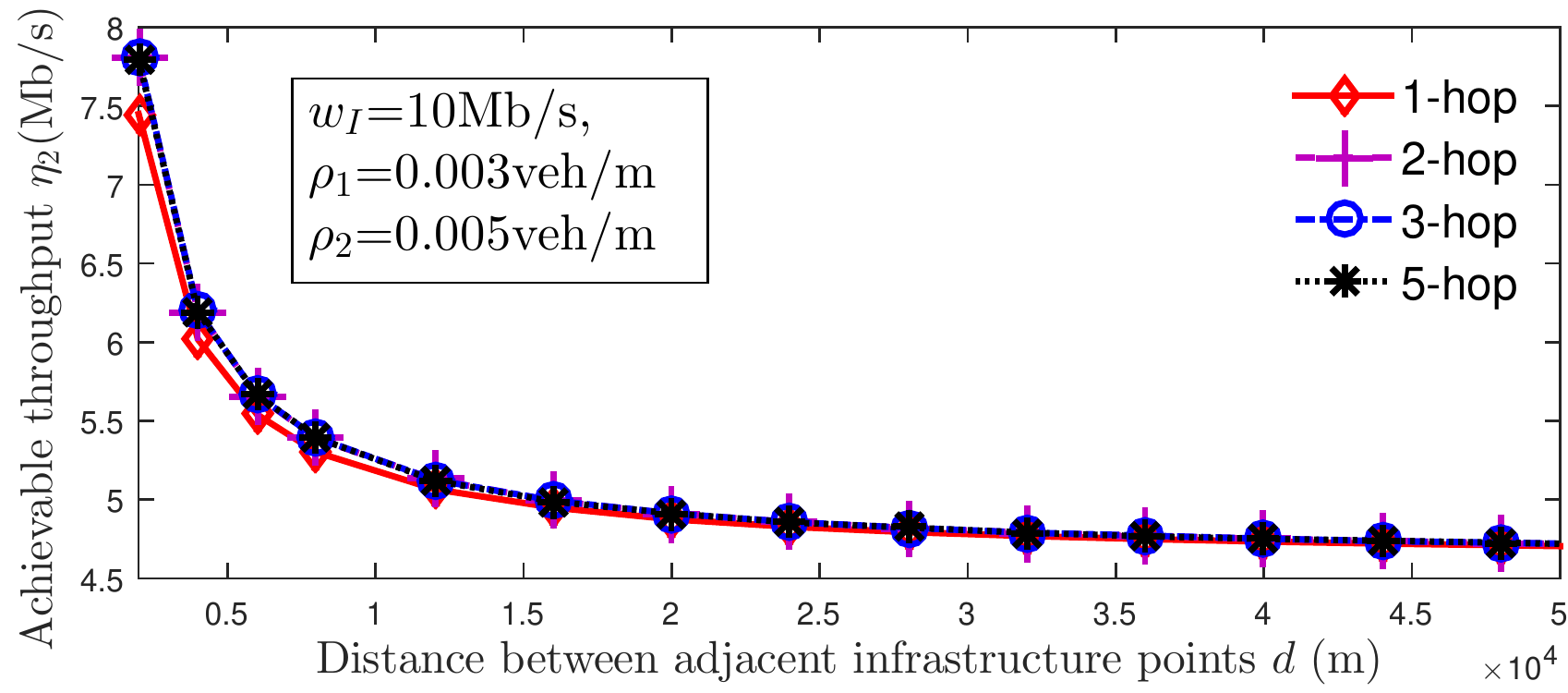} \caption{A comparison between throughput achieved when allowing one-hop communication
and multi-hop communications.}
\label{K-hop} 
\end{figure}

\textcolor{black}{Fig. \ref{K-hop} compares the throughput achieved
by allowing only one-hop communication and allowing both $k$-hop
($k=2,3,5)$ V2I communications between the VoI and infrastructure
and $k$-hop V2V communication between the VoI and }\textit{\textcolor{black}{helpers}}\textcolor{black}{.
It is shown that allowing multi-hop communications beyond one hop
has little impact on the throughput. Particularly, as pointed out
in the end of Section \ref{subsec:Wireless-Communication-Model},
in our considered scenario, allowing multi-hop V2V communication only
helps to balance the distribution of information among }\textit{\textcolor{black}{helpers}}\textcolor{black}{{}
but do not increase the net amount of information available in the
network. The marginal increase in the achievable throughput comes
from multi-hop V2I communications between the VoI and infrastructure,
because it allows the VoI having longer connection time with the infrastructure.}

\begin{figure}
\begin{centering}
\includegraphics[width=8cm]{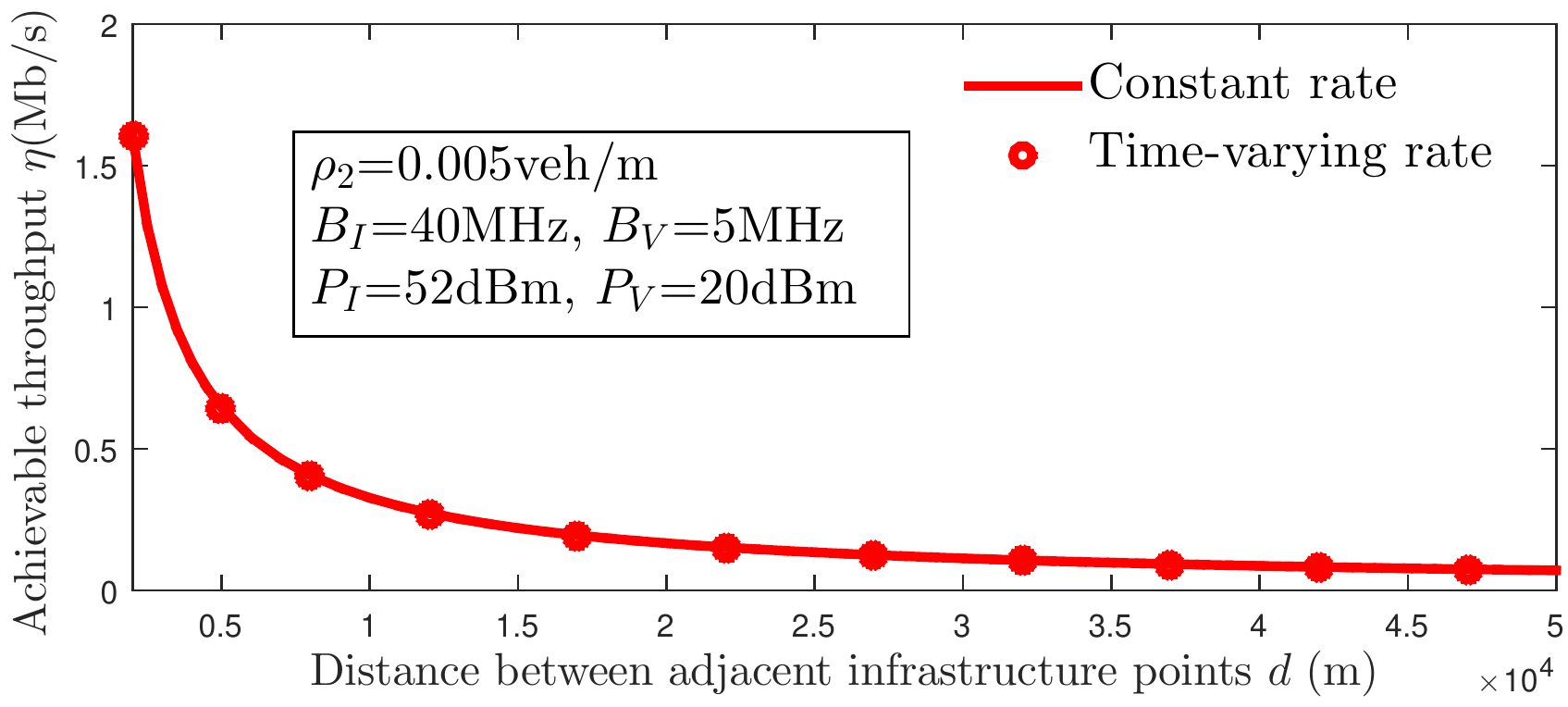}
\par\end{centering}
\caption{A comparison between throughput achieved from constant channel model
and time-varying channel model which considering Rayleigh fading and
path loss.}

\label{Path_loss}

\end{figure}

Fig. \ref{Path_loss} compares throughput achieved from the constant
channel model with that from the time-varying channel model, and shows
that our analysis under the constant channel model is applicable to
a more realistic time-varying channel model which considers both fading
and path loss. Specifically, for the time-varying channel model, we
adopt the model used in \cite{Wang2013} that considers Rayleigh fading
and path loss, from which the transmission rate is given by $w_{I}^{'}=B_{I}\log_{2}\left(1+P_{I}|\beta d_{i}^{-2}|^{2}\right)$
and $w_{V}^{'}=B_{V}\log_{2}\left(1+P_{V}|\beta d_{ij}^{-2}|^{2}\right)$,
with the bandwidth and transmit power of each infrastructure and vehicle
being $B_{I}$=40MHz, $P_{I}$=52dBm and $B_{V}$=5MHz, $P_{V}$=20dBm
\cite{Zheng13} respectively. Parameter $\beta$ is the Gaussian random
variable with mean 0 and variance 1 and $d_{i}$, $d_{ij}$ are the
distances between a vehicle and its associated infrastructure point,
between vehicle and vehicle when conducting V2I and V2V communications
respectively. The above settings of $B_{I}$, $P_{I}$, $B_{V}$ and
$P_{V}$ implies that the network is in the V2V-Limited regime. By
dividing the total coverage length of the transmitter (infrastructure
or vehicle) into $K$ (here we set $K$=1000) small segments, the
average channel throughput $w_{I}$ and $w_{V}$ in the time-varying
channel model can be obtained by averaging the transmission rates
of all segments. This obtained average throughput $w_{I}$ and $w_{V}$
are then used in our constant channel model. It is obvious from Fig.
\ref{Path_loss} that the achievable throughput from the above two
channel models match each other. This phenomenon can be explained
by equation \eqref{eta2} which shows that the achievable throughput
in V2V-Limited regime is a linear function of $w_{I}$ and $w_{V}$.
Then it follows that $E[\eta(w_{I},w_{V})]=\eta(E[w_{I}],E[w_{V}])$,
which implies that for time-varying channels, the time-varying values
of $w_{I}^{'}$ and $w_{V}^{'}$ can be replaced by the respective
time-averaged throughput of V2I and V2V communications and our analysis
still applies.

\section{Conclusions \label{sec:Conclusion-and-Future}}

This paper proposed a cooperative communication strategy for vehicular
networks with a finite vehicular density by utilizing V2I communications,
V2V communications, mobility of vehicles, and cooperations among vehicles
and infrastructure to improve the throughput. A detailed analysis
for the achievable throughput was presented and the closed-form expression
of achievable throughput (or its upper and lower bound) was obtained
in three different regimes we classified in our analysis based on
the relationship between the data rates of V2I communications, V2V
communications, and the speeds of vehicles. Numerical and simulation
results show that the proposed cooperative strategy can significantly
improve the achievable throughput of vehicular networks even when
traffic density is rather low. Simulation results show that our analysis
can be extended to more realistic models such as time-varying speed
model, log-normal shadowing model and time-varying channel model considering
fading and path loss. Our results shed insight on the optimum design
of vehicular network infrastructure and the design of optimum cooperative
communication strategies in finite vehicular networks to maximize
the throughput. 

\begin{IEEEbiography}{Jieqiong Chen}
(S'16) received the Bachelor\textquoteright s degree in Engineering
from Zhejiang University, Zhejiang, China, in 2012, and she is currently
working toward the Ph.D. degree in engineering at the University of
Technology Sydney, Sydney, Australia. Her research interests in the
area of vehicular networks.
\end{IEEEbiography}

\begin{IEEEbiography}{Guoqiang Mao}
(S'98-M'02-SM'07) received PhD in telecommunications engineering
in 2002 from Edith Cowan University. Between 2002 and 2013, he was
an Associate Professor at the School of Electrical and Information
Engineering, the University of Sydney. He currently holds the position
of Professor of Wireless Networking, Director of Center for Real-time
Information Networks at the University of Technology, Sydney. He has
published more than 100 papers in international conferences and journals,
which have been cited more than 3000 times. His research interest
includes intelligent transport systems, applied graph theory and its
applications in networking, wireless multi-hop networks, wireless
localization techniques and network performance analysis. He is an
Editor of IEEE Transactions on Vehicular Technology and IEEE Transactions
on Wireless Communications, and a co-chair of IEEE Intelligent Transport
Systems Society Technical Committee on Communication Networks. 
\end{IEEEbiography}

\begin{IEEEbiography}{Changle Li}
(M'09) received the B.E. degree in microwave telecommunication engineering
and the M.E. and Ph.D. degrees in communication and information system
from Xidian University, Xi?an, China, in 1998, 2001, and 2005, respectively.
From 2006 to 2007, he was with the Department of Computer Science,
University of Moncton, Moncton, NB, Canada, as a Postdoctoral Researcher.
From 2007 to 2009, he was an expert researcher at the National Institute
of Information and Communications Technology, Japan. He is currently
a Professor with the State Key Laboratory of Integrated Services Networks,
Xidian University. His research interests include intelligent transportation
systems, vehicular networks, mobile ad hoc networks, and wireless
sensor networks. 
\end{IEEEbiography}

\begin{IEEEbiography}{Ammar Zafar}
obtained his PhD in Wireless Communications from King Abdullah University
of Science and Technology in 2014. In November 2014, he joined Center
of Real-time Information Networks at the University of Technology
Sydney as a postdoctoral research fellow. Ammar Zafar's research interests
include vehicular networks, multi-user scheduling, wireless communication
theory and intelligent transportation systems. 
\end{IEEEbiography}

\begin{IEEEbiography}{Albert Y. Zomaya}
is currently the chair professor of high-performance computing and
networking and Australian Research Council professorial fellow in
the School of Information Technologies, The University of Sydney,
Sydney, Australia. He is also the director of the Centre for distributed
and high-performance computing which was established in late 2009.
He is the author/co-author of seven books, more than 370 papers, and
the editor of nine books and 11 conference proceedings. He is the
editor in chief of the IEEE Transactions on Computers and serves as
an associate editor for 19 leading journals. He is the recipient of
the Meritorious Service Award in 2000 and the Golden Core Recognition
in 2006, both from the IEEE Computer Society. He is a chartered engineer
(CEng), a fellow of the AAAS, the IEEE, the IET (UK), and a distinguished
engineer of the ACM.
\end{IEEEbiography}

\end{document}